\documentclass[a4paper,11pt]{article}
\usepackage[margin=0.5in]{geometry} 
\usepackage{enumerate}   
\usepackage{jcappub}
\usepackage{mathtools} 

\usepackage{amsmath}
\usepackage{amsthm}
\usepackage{amssymb}
\usepackage{natbib}
\usepackage{aas_macros}
\usepackage{xspace}
\usepackage[utf8]{inputenc}
\usepackage[T1]{fontenc}
\usepackage{hyperref}
\usepackage{caption}
\usepackage{verbatim}

\hypersetup{
	unicode,
	pdfauthor={Author One, Author Two, Author Three},
	pdftitle={Growth history and quasar bias evolution at $z<3$ from Quaia},
	pdfsubject={Growth history and quasar bias evolution at $z<3$ from Quaia},
	pdfproducer={LaTeX},
	pdfcreator={pdflatex}
}

\usepackage{graphicx, color}

\newcommand{\planck}{{\sl Planck}\xspace}
\newcommand{\quaia}{\textit{Quaia}\xspace}
\newcommand{\gaia}{{\sl Gaia}\xspace}
\newcommand{\unwise}{{\sl unWISE}\xspace}
\newcommand{\nv}{\hat{\bf n}}
\newcommand{\wtj}[6]{\left(\begin{array}{ccc} #1 & #2 & #3\\#4 & #5 & #6\end{array} \right)}

\definecolor{darkgreen}{RGB}{0, 153,0}
\definecolor{darkblue}{RGB}{0, 0,153}

\newcommand{\lmax}{\ell_{\rm max}}

\newcommand{\newv}[1]{\textcolor{black}{#1}\xspace}

\title{\boldmath Growth history and quasar bias evolution at $z<3$ from Quaia}

\author[a,b,c]{Giulia Piccirilli,}
\author[d, e]{Giulio Fabbian,}
\author[f]{David Alonso,}
\author[g]{Kate Storey-Fisher,}
\author[h]{Julien Carron,}
\author[i]{Antony Lewis,}
\author[f]{Carlos {Garc{\'\i}a-Garc{\'\i}a}}

\affiliation[a]{Dipartimento di Fisica, Università di Roma Tor Vergata, via della Ricerca Scientifica, 1, 00133, Roma, Italy}
\affiliation[b] {INFN - Sezione di Roma 2, Università di Roma Tor Vergata, via della Ricerca Scientifica, 1, 00133 Roma, Italy}
\affiliation[c]{Dipartimento di Fisica, Università La Sapienza, P.le A. Moro 2, 00185, Roma, Italy}
\affiliation[d]{Center for Computational Astrophysics, Flatiron Institute, 162 Fifth Avenue, New York, NY, 10010, USA}
\affiliation[e]{School of Physics and Astronomy, Cardiff University, The Parade, Cardiff, Wales CF24 3AA, United Kingdom}
\affiliation[f]{Department of Physics, University of Oxford, Denys Wilkinson Building, Keble Road, Oxford OX1 3RH, United Kingdom}
\affiliation[g]{Donostia International Physics Center, Manuel Lardizabal Ibilbidea, 4, 20018 Donostia, Gipuzkoa, Spain}
\affiliation[h]{{Universit{\' e}} de Genève, D{\' e}partement de Physique Th{\' e}orique et CAP, 24 Quai Ansermet, CH-1211 Genève 4, Switzerland }
\affiliation[i]{Department of Physics \& Astronomy, University of Sussex, Brighton BN1 9QH, UK}

\emailAdd{giulia.piccirilli@roma2.infn.it}

\abstract{We make use of the \gaia-\unwise quasar catalogue, \quaia, to constrain the growth history out to high redshifts from the clustering of quasars and their cross-correlation with maps of the Cosmic Microwave Background (CMB) lensing convergence. Considering three tomographic bins, centred at redshifts $\bar{z}_i = [0.69, 1.59, 2.72]$, we reconstruct the evolution of the amplitude of matter fluctuations $\sigma_8(z)$ over the last $\sim12$ billion years of cosmic history. In particular, we make one of the highest-redshift measurements of $\sigma_8$ ($\sigma_8(z=2.72)=0.22\pm 0.06$), finding it to be in good agreement (at the $\sim1\sigma$ level) with the value predicted by $\Lambda$CDM using CMB data from \planck. We also used the data to study the evolution of the linear quasar bias for this sample, finding values similar to those of other quasar samples, although with a less steep evolution at high redshifts. \newv{Finally, we study the potential impact of foreground contamination in the CMB lensing maps and, although we find evidence of contamination in cross-correlations at $z\sim1.7$ we are not able to clearly pinpoint its origin as being Galactic or extragalactic. Nevertheless, we determine that the impact of this contamination on our results is negligible.}}
\begin{document}
\maketitle
\flushbottom

\section{Introduction}\label{sec:introduction}
  The evolution of the amplitude of density fluctuations across cosmic time is a key cosmological observable, sensitive to the energy content of the Universe, as well as the laws of gravity that govern the growth of structure. The increase in the quantity and variety of cosmological datasets in the last two decades has provided us with a wide range of probes able to reconstruct this growth \cite{Huterer:2022dds}. On the one hand, probes sensitive to the peculiar velocity field, such as redshift-space distortions \citep{BOSS:2016wmc,Hou:2020rse} or velocity surveys \citep{Stahl:2021mat}, are able to measure the growth {\sl rate} with remarkable precision, constraining the speed at which the structure grows over time. On the other hand, gravitational lensing directly probes the amplitude of matter inhomogeneities. For instance, cosmic shear data and the lensing of the Cosmic Microwave Background (CMB) are sensitive to the cumulative distribution of matter along the line of sight from the source, and thus constrain the amplitude of inhomogeneities at intermediate redshifts, providing also some information about the evolution of these fluctuations, if sources at different redshifts are combined. The combined analysis of galaxy clustering and cosmic shear data (the now commonplace ``3$\times$2-point'' analysis \cite{Giri_2023}), provides additional sensitivity to the redshift dependence of growth. This is possible by exploiting the fact that the galaxy overdensity largely probes matter structures locally, and thus cross-correlations of cosmic shear with galaxies at different redshifts can be used to reconstruct the growth history with finer granularity. The technique is particularly powerful when employing CMB lensing (a methodology now commonly labelled ``CMB lensing tomography'' \citep{Giannantonio_2016, Peacock_2018}). With a lensing source (i.e. the CMB) located behind any other tracer of the large-scale structure (LSS), CMB lensing tomography can be used to reconstruct the growth history up to high redshifts.
  
  Previous attempts at reconstructing the growth history, using redshift-space distortion, $3\times2$ point data, and/or CMB lensing tomography, have revealed hints of a potentially lower growth at low redshifts with respect to the predictions from $\Lambda$CDM favoured by primary CMB data from e.g. \planck \citep{Heymans_2021, Krolewski_2020, Omori_2018, Nguyen:2023fip}. These outcomes align with the evidence for the so-called $S_8$ tension, the mild disagreement in the value of $S_8\equiv\sigma_8\sqrt{\Omega_M/0.3}$ between low-redshift LSS data and CMB experiments. Since this result is not borne out by constraints based on the CMB lensing power spectrum in combination with Baryon Acoustic Oscillations (BAO) data \citep{Carron_2022,ACT:2023kun}, it is therefore unclear whether the source of this tension is a potential astrophysical systematic affecting low-redshift tracers on small scales. Alternative measurements of growth at high redshifts are, therefore, highly valuable to pin down the origin of this tension. However, previous growth reconstruction efforts have been somewhat limited by the lack of local LSS tracers at high redshifts ($z\gtrsim1$, \citep{Doux_2022, Hamana_2020,Hamana_2020_erratum}). Although infrared data from, e.g. \unwise are able to reach redshifts $z\sim1.5$ \citep{unWiseKrolewski,unWiseFarren}, the $z\gtrsim2$ regime is largely unexplored. Radio continuum galaxy surveys are sensitive to this regime, but the complexity of obtaining a reliable redshift distribution for high-density continuum samples makes it difficult to exploit them for precision cosmology \cite{lofar1,lofar2}. Other probes of the high-redshift LSS will become available with next-generation surveys, such as large Lyman-break galaxy samples (LBGs -- see \cite{Miyatake_2022} for the first cosmological analysis of an LBG sample), or intensity mapping \cite{Yohana_2019}. Until these data are available at scale, perhaps our best avenue to explore the high-redshift Universe are quasar catalogues.

  Quasars, together with the Lyman-$\alpha$ forest, have provided our highest-redshift measurements of the redshift-distance relation via BAO \cite{Neveux_2020}, and they have been exploited to constrain large-scale cosmological observables, such as primordial non-Gaussianity, due to the enormous volumes they are able to probe. Recently, \cite{StoreyFisher_2023} presented \quaia, a quasar catalogue that covers almost the entire celestial sphere, constructed by combining the \gaia catalogue of quasar candidates with infrared photometry from \unwise. The availability of accurate and precise spectrophotometric redshifts allowed \cite{Alonso_2023} to derive relatively tight constraints on $\Lambda$CDM from the auto-correlation of \quaia and its cross-correlation with CMB lensing maps from \planck. Our aim in this paper is to extend the analysis of \cite{Alonso_2023}, which focused on $\Lambda$CDM constraints, considering only two broad redshift bins. Here, we will take advantage of the wide range of redshifts covered by \quaia to reconstruct the growth history in a model-independent way out to $z\sim3$, beyond the range explored by previous analyses (e.g. \cite{GG_2021,unWiseKrolewski,DESs8z,WhiteLRG,unWiseFarren}). In doing so, we will also study the redshift evolution of the quasar bias in the \quaia sample and attempt to isolate potential sources of systematic contamination in the CMB lensing maps, first identified in \cite{Alonso_2023}.

  The paper is structured as follows. Section \ref{sec:data} presents the \quaia catalogue and the various CMB lensing maps used in our analysis. The theoretical model used in this study, as well as the various data analysis and parameter inference methods employed here, are described in Section \ref{sec:methods}. The results of our analysis regarding growth and quasar bias evolution, as well as CMB lensing systematics, are presented in Section \ref{sec:results}. We then summarise our findings and conclude in Section \ref{sec:conclusions}.

\section{Data}\label{sec:data}
  \subsection{\quaia}\label{ssec:data.quaia}
  
    The \gaia-\unwise Quasar catalogue (\quaia hereafter) is a quasar sample that covers the entire sky and contains almost 1.3 million sources with magnitude $G<20.5$. The catalogue, described in detail in \cite{StoreyFisher_2023}, was obtained by combining the sample of quasar candidates in the third \gaia data release \cite{GaiaDR3_2016} together with infrared photometry from the \unwise reprocessing of the Wide-Field Infrared Survey Explorer (WISE) \cite{Meisner_2019} to improve the purity of the sample and the quality of the redshift estimates. \quaia sources are assigned a spectrophotometric redshift estimated using a $k$-nearest neighbours algorithm that combines the \gaia-estimated spectroscopic redshift and photometry from both \gaia and \unwise. The algorithm is trained on spectroscopic data from the Sloan Digital Sky Survey DR16Q \cite{Lyke_2020}. Note that we use the latest version of the \quaia catalogue\footnote{Publicly available at \url{https://zenodo.org/records/8060755}.}, which is slightly different from that used in \cite{Alonso_2023}. The latest version uses an updated dust map, which improves the redshift uncertainties slightly; it also includes an updated selection function taking into account \unwise survey properties, as well as improved modelling of high-completeness regions.
\begin{figure}[h!]
    \centering
    \includegraphics[width = 0.75\linewidth]{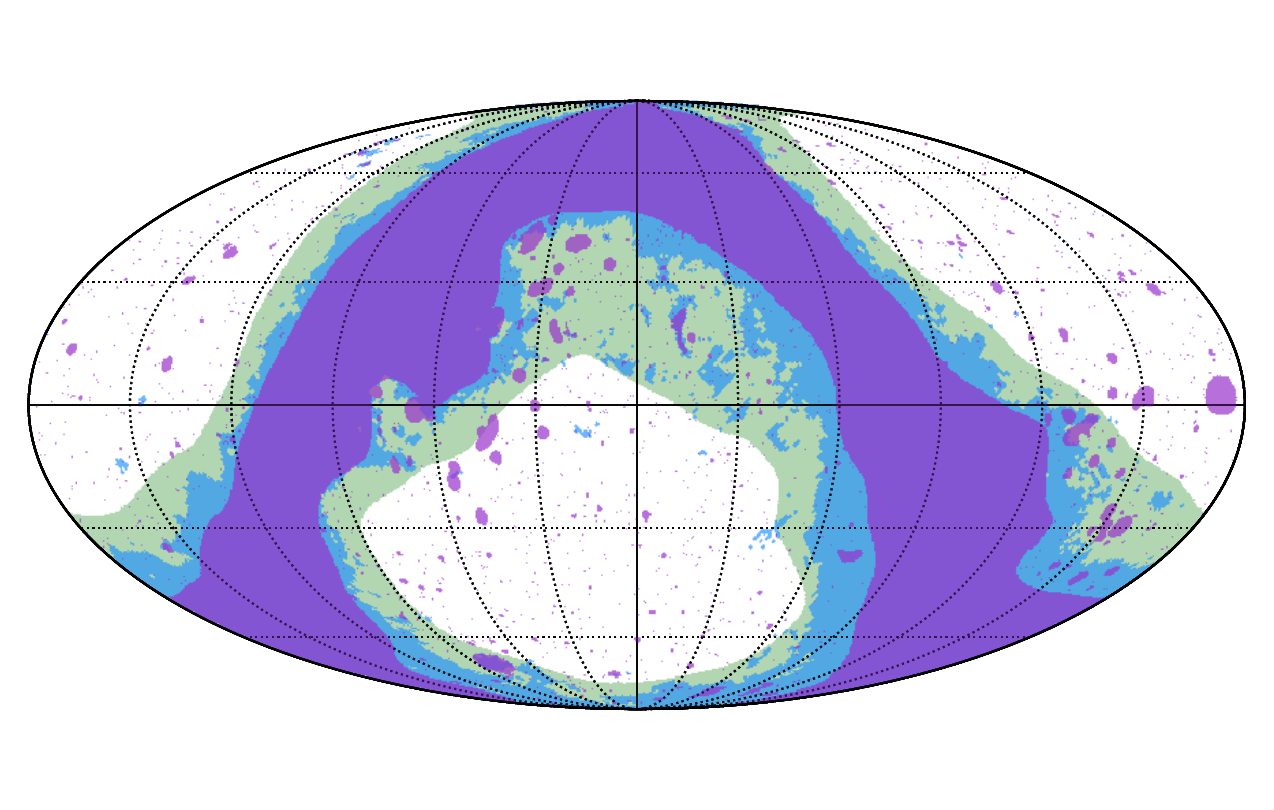}
    \caption{Mollweide projection in Equatorial coordinates of the masked area of the CMB lensing map (purple), the first \quaia redshift bin (blue), and of the $40\%$ Galactic mask made available by \planck (green).} \label{fig:footprints}
  \end{figure}
    In our analysis, we divided the total redshift distribution of the sources in three redshift bins, defined by the bin boundaries $z_Q<1.0$, $1.0<z_Q<2.3$, and $z_Q>2.3$, where $z_Q$ is the \quaia spectro-photometric redshift estimate. The resulting bins are centred at $\bar{z}_i = [0.69, 1.59, 2.7]$.
    This provides us with a better handle on growth evolution than the two bins used in \cite{Alonso_2023}, allowing us to also isolate the signal from the structure at $z\sim3$. The angular selection function for each of these bins was generated using the methodology described in \cite{StoreyFisher_2023}. We generate a projected ovedensity map {\tt HEALPix}\footnote{\url{https://healpix.sourceforge.io/}} \cite{Gorski_2005} for each redshift bin, defined as
    \begin{equation}
      \delta_g(\nv) = \frac{N(\nv)}{\bar{N}w(\nv)} - 1,
    \end{equation}
    where $\hat{{\bf n}}$ is the unit vector in the pixel direction, $N(\hat{{\bf n}})$ is the number of sources in the pixel, $w(\hat{{\bf n}})$ is the value of the selection function in the pixel, and $\bar{N}$ the mean number of quasars per pixel, computed as:
    \begin{equation}\label{eq:over_map}
      \bar{N} = \frac{\sum_{\nv} N(\nv)}{\sum_{\nv} w(\nv)}.
    \end{equation}
    We mask all areas of the sky in which the selection function takes values $w<0.5$. This cut leaves $57\%, 59\%$ and $38\%$ available in each redshift bin. The sky mask for the first redshift bin is shown in Figure~\ref{fig:footprints}.

\begin{figure}
    \centering
    \includegraphics[width = 0.75\linewidth]{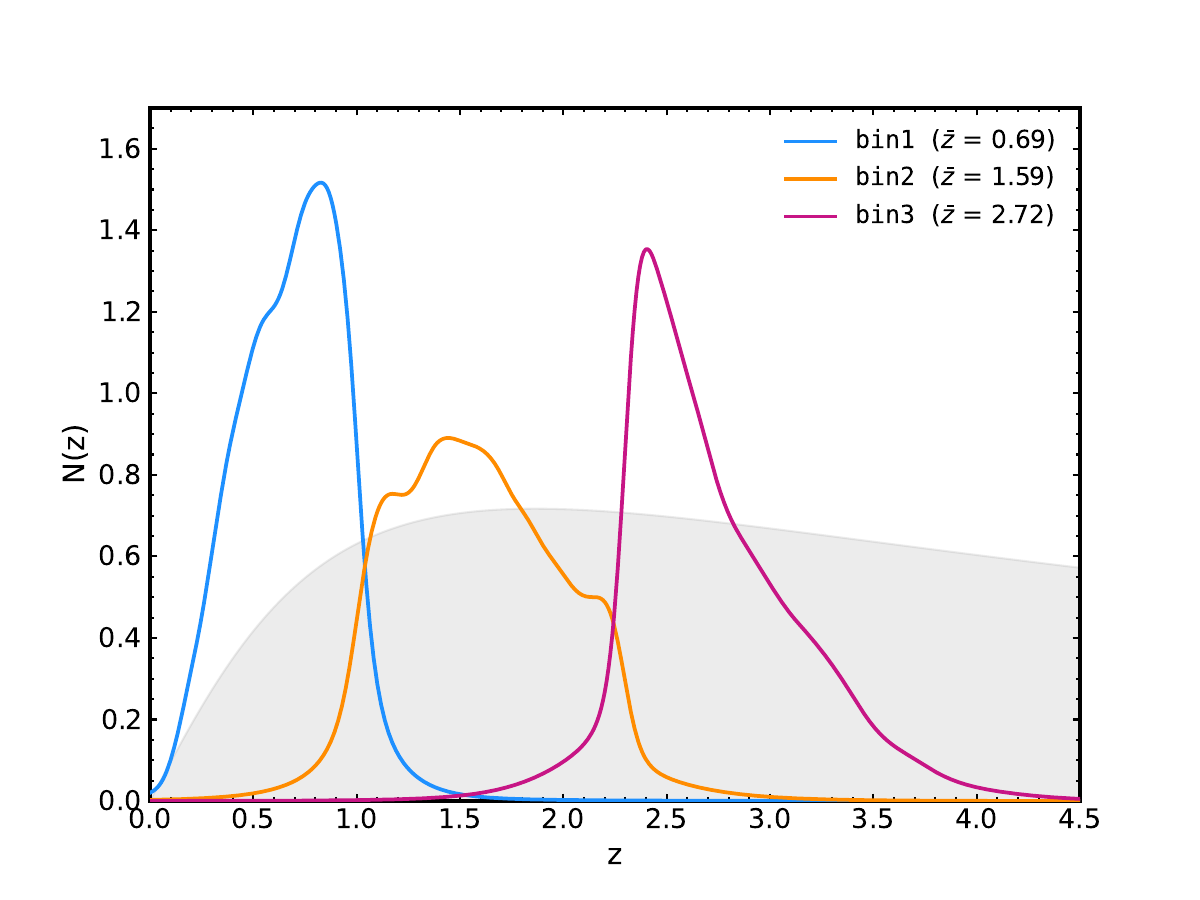}
    \caption{Normalized redshift distribution of the three \quaia redshift bins, defined in Section~\ref{ssec:data.quaia}. The grey-shaded area represents the CMB lensing kernel $W_{\kappa}(z)$, with an arbitrary normalisation for visualisation purposes.}
    \label{fig:kg_kernels}.
  \end{figure}

    The redshift distribution for each bin was calculated by stacking the individual redshift probability density functions (PDFs) of each source (parameterized as normal distributions determined by the estimated redshift and its uncertainty) in the bin. This was found in \cite{Alonso_2023} to be a reasonably accurate estimate of the redshift distribution when compared with direct calibration estimates. The normalised redshift distributions of the three bins are shown in Figure~\ref{fig:kg_kernels}, together with the CMB lensing kernel (see Section~\ref{ssec:methods.theoretical_model}).

  \subsection{CMB lensing maps}\label{ssec:data.cmb}
    We use the CMB lensing data from the \planck satellite that measured the angular power spectrum of the CMB lensing convergence on a range of multipoles $8\leq \ell \leq 2048$ \citep{Planck_2020b, Planck_len2020}, 
    detecting the signal at $40\sigma$ significance from the minimum variance estimator combining temperature and polarization information. Ref.~\cite{Alonso_2023} observed hints of potential residual systematics in the \planck CMB lensing maps when cross-correlated with \quaia sources at $z\sim2$. Contamination from the Cosmic Infrared Background (CIB) is an attractive candidate to explain this effect (since it appears in a cross-correlation and is mostly evident at high redshifts), but a more thorough analysis is needed to fully determine its origin and statistical significance. To explore this in detail, we have made use of eight different convergence maps reconstructed from the \planck data, resulting from different data combinations and lens reconstruction techniques. In particular, the maps were obtained from Planck PR4\footnote{\url{https://github.com/carronj/planck_PR4_lensing}} data generated through the {\tt NPIPE} pipeline \cite{Carron_2022}. The maps considered are the following.
    \begin{enumerate}[i.)]
      \item {\bf Generalized Minimum-Variance (``{\sf GMV}'') map}. Based on PR4 data, this lensing reconstruction has an improved signal-to-noise ratio of up to $\sim 20\%$ with respect to the previous \planck data release (PR3\footnote{\url{https://wiki.cosmos.esa.int/planck-legacy-archive/index.php/Lensing}}) results due to a joint inverse-variance Wiener filtering of the CMB temperature and polarisation maps accounting for inhomogeneous noise with an optimal weighting.
      \item {\bf Polarisation-only map (``{\sf Pol-only}'') map}, constructed through a quadratic estimator employing only polarisation data, and hence largely immune to contamination from unpolarized Galactic or extragalactic foregrounds, such as the CIB or the Sunyaev-Zeldovich effect.
      \item {\bf No-temperature (``{\sf No-TT}'') map}. This corresponds to a minimum-variance quadratic estimator that down-weights $TT$ correlations but preserves $TE$ information. Specifically, this map is built in the same way described in the \planck~PR3 paper~\cite{Planck_len2020} (with the PR4 CMB maps), but leaving out the $TT$-block in the calculation of the minimum-variance combination. This should still be robust to unpolarized foregrounds while preserving more sensitivity than {\sf Pol-only}.
      \item {\bf Temperature alone (``{\sf TT-only}'') map}. A quadratic estimator that uses only temperature information.
      \item \newv{{\bf {\sf TE-only} map}: a CMB reconstruction making use only of the correlation between temperature and $E$-mode polarization, in order to establish whether any differences observed between the {\sf GMV}/{\sf TT} maps, and the {\sf Pol-only} and {\sf No-TT} maps are driven only by the polarization data.}
      \item {\bf SZ-deprojected map (``{\sf TT-noSZ}'')}. A temperature-only map based on the {\tt SMICA} foreground-cleaned CMB maps, where the thermal Sunyaev-Zeldovich effect is deprojected \cite{Planck_len2020}. The response of this map to CIB contamination should be different from the {\sf GMV} or TT-only maps (potentially suffering more from it, since the frequency weighting prioritises removing SZ at the expense of increasing contamination from any other component).
      \item \newv{{\bf Minimum-variance, source-hardened map (``{\sf MVh}'')}: Bias-hardening \citep{Namikawa_2013, Osborne_2014} effectively nulls out the contribution to the reconstructed $\kappa$ map from any contaminant with a known shape of its power spectrum. Source-hardening corresponds to the application of bias-hardening to the case of Poisson-sampled point sources. The resulting $\kappa$ map is therefore largely immune to any unclustered point-source-like contaminant. Here, the minimum variance consists of an inverse variance weighted combination of all the different reconstruction channels, where the CMB temperature and polarisation maps are filtered separately prior to lensing reconstruction. The bias-hardening operation was performed only for the temperature estimator.}
      \item \newv{{\bf Temperature-only, source-hardened map (``{\sf TTh}'')}: as {\sf TT-only} but applying source hardening.}
   \end{enumerate}
   The processing of all CMB lensing maps involves a standardised procedure. First, the spherical harmonic coefficients are rotated to the equatorial coordinates and then transformed into a HEALPix map with a resolution parameter of $N_{\rm side}= 512$. To avoid aliasing, all multipoles with $\ell>3 N_{\rm side}$ are filtered out before creating the map. The same angular mask, publicly provided by \planck and shown in Figure~\ref{fig:footprints}, is applied to all CMB lensing maps after an analogous rotation and degradation. This mask excludes areas susceptible to contamination by Galactic foregrounds, and removes SZ-clusters detected at high significance as well as a series of bright extragalactic sources, leaving approximately 67\% of the total sky.

   \newv{Additionally, in order to test for potential contamination from Galactic foregrounds (either in the $\kappa$ maps or in \quaia), we will also make use of the 40\% Galactic mask made available by \planck. Results using this mask will be labelled {\sf Gal. mask}.}

\section{Methods}\label{sec:methods}
  \subsection{Theoretical model}\label{ssec:methods.theoretical_model}
    In our analysis, we consider two projected probes: the angular overdensity of \quaia quasars $\delta_g(\nv)$, and the CMB lensing convergence $\kappa(\nv)$. 
    
    On the one hand, the quasars are biased tracers of the underlying matter density field, $\delta_m({\bf x})$. Assuming a linear bias model (valid on the scales used in this analysis), $\delta_g(\nv)$ is related to $\delta_m({\bf x})$ via
    \begin{equation}
      \delta_g(\nv) = \int dz\,b(z)\,N(z)\,\delta_m(\chi(z)\nv),
    \end{equation}
    where $N(z)$ is the normalized redshift distribution of the quasars, and $b(z)$ is the linear bias. 
    
    On the other hand, CMB lensing is related to the distortion of the CMB photon trajectories due to the presence of the gravitational potential of the LSS of the Universe, and hence it is an unbiased tracer of the matter fluctuations that source this gravitational potential. The relation between $\kappa(\nv)$ and $\delta_m({\bf x})$, assuming $\Lambda$CDM, is:
    \begin{equation}
      \kappa(\nv) = \int dz\,\frac{3}{2}\Omega_{m,0}H_0^2\frac{1+z}{H(z)}\chi(z)\left(1-\frac{\chi(z)}{\chi_*}\right) \delta_m(\chi(z)\nv),
    \end{equation}
    where $\Omega_{m,0}$ is the matter density today, $H(z)$ is the Hubble parameter, and $\chi_*$ is the comoving distance to the last-scattering surface at $z_*\simeq 1100$.
    
    Now, let $\ X(\nv)$ be a projected field related to the matter fluctuations through the kernel $W_{\rm x}(\chi)$ via
    \begin{equation}
      \ X(\nv)=\int dz\,W_{\rm x}(\chi(z))\,\delta_m(\chi(z)\nv).
    \end{equation}
    The angular power spectrum between two such quantities, $\ X$ and $\ Y$, is related to the matter power spectrum $P(k,z)$ via 
    \begin{equation}\label{eq:kg_th}
        C_{\ell}^{\rm XY} = \int dz \frac{H(z)}{c\chi^2(z)} W_{\rm x}(z)W_{\rm y}(z)P\left(k = \frac{(\ell+1/2)}{\chi(z)}, z\right).
    \end{equation}
    Note that the above expression holds for the Limber approximation \cite{Limber_53}, valid on the scales used in this analysis. The power spectrum in Equation~\ref{eq:kg_th} depends upon the overlap between the radial kernels for the quasars overdensity and the lensing convergence, which, based on the above discussion, are:
    \begin{equation}\label{eq:kernel}
        W_g(z) = b(z)\,N(z),\hspace{12pt}
        W_{\kappa}(z) = \frac{3\Omega_{m,0} H_0^2}{2} \frac{1+z}{H(z)}\chi(z)\left(1-\frac{\chi(z)}{\chi_*}\right).
    \end{equation}
    
    Our analysis will make use of the galaxy auto-correlation, $C_\ell^{gg}$, and its cross-correlation with CMB lensing, $C_\ell^{\kappa g}$. To compute these spectra, we need four main ingredients: a cosmological model (e.g. the $\Lambda$CDM model), a non-linear model for the matter power spectrum $P(k,z)$ for which we will use the {\tt HaloFit} model as described in \cite{Takahashi_2012}, the redshift distribution $N(z)$, and a model for galaxy bias $b(z)$. As discussed in the previous section, we calculate the redshift distributions by stacking the individual redshift PDFs of all sources in each bin, which was found to be sufficiently accurate in \cite{Alonso_2023}. 
    
    We assume the redshift evolution of $b(z)$ to be the same in all three redshift bins, although with potentially different amplitudes. Specifically, we assume that the redshift dependence of the bias within each redshift bin is given by the fitting function of \cite{Laurent_2017}, and thus the bias in bin $i$ is given by:
    \begin{equation}\label{eq:bias}
      b^i(z) = b_g^i[0.278((1+z)^2-6.565)+2.393],
    \end{equation}
    where $b_g^i$ is a free parameter of the model that controls the amplitude of $b(z)$. The specific redshift dependence of Equation~\ref{eq:bias} was obtained by \cite{Laurent_2017} by fitting a quadratic polynomial to the auto-correlation of eBOSS quasars at different redshifts. The best-fit of \cite{Laurent_2017} would be recovered for $b_g^i=1$. Note that the specific form of the redshift dependence is only relevant within each of the redshift bins explored, since we allow for a different overall amplitude in each bin. The detailed form of the bias evolution was found by \cite{Alonso_2023} to have little impact on the final $\Lambda$CDM constraints.
    
    As described in \cite{Alonso_2023}, magnification bias has a negligible effect for \quaia, since the slope of the cumulative flux distribution is close to $s=0.4$. We nevertheless account for the impact of this effect in our analysis.
    
  \subsection{Growth evolution}\label{ssec:methods.growth_evolution}
    The main aim of this paper is to use \quaia to reconstruct the growth history, parametrized in terms of $\sigma_8(z)$, in a model-independent way. This is particularly interesting at high redshifts ($z\gtrsim1.5$) where the availability of LSS probes is limited, and which \quaia is ideally suited to explore. 
     
    Within $\Lambda$CDM, and neglecting the presence of massive neutrinos (i.e. $\sum_{\nu} m_{\nu}=0$), the time dependence of the linear matter power spectrum is factorizable as:
    \begin{equation}\label{eq:lingrowth}
        P_{\rm lin}(k,z) = D(z)^2P_{\rm lin}(k, z=0),
    \end{equation}
    where $D(z)$ is the linear growth factor. Here, we will parametrize deviations from the $\Lambda$CDM prediction for $D(z)$ defining the growth factor as:
    \begin{equation}\label{eq:growthpar}
      D(z)=D_{\rm fid}(z)\,[1+\Delta(z)],
    \end{equation}
    where $D_{\rm fid}(z)$ is the linear growth factor for a fixed fiducial cosmology, and $\Delta(z)$ is a linear interpolation of the three free parameters $\Delta(\bar{z})_i\,i\in\{1,2,3\}$, corresponding to the value of $\Delta(z)$ at the mean redshift of the three redshift bins. Note that, when reconstructing $D(z)$, we do not enforce the normalization condition $D(z=0)=1$, and thus $\Delta(z)$ parametrizes both deviations in the redshift dependence of the growth factor and in the overall amplitude of the linear matter power spectrum at $z=0$. For this reason, in this case, we fix the value of $\sigma_8$ to $\sigma_8^{\rm fid}=0.8102$ (the best-fit value found by \planck) when calculating $P_{\rm lin}(k,z=0)$. We will present our results in terms of the redshift-dependent $\sigma_8(z)$, which we calculate as $\sigma_8(z)=\sigma_8^{\rm fid}\,D(z)$, with $D(z)$ given by Equation~\ref{eq:growthpar}.

  \subsection{Angular power spectra}\label{ssec:methods.power_spectra}
    The auto and cross correlation spectra used in the analysis, are computed through the pseudo-$C_{\ell}$ formalism \citep{MASTER} as implemented in \texttt{NaMaster}\footnote{\url{https://github.com/LSSTDESC/NaMaster}} \cite{Alonso_18}, which we describe below. Let us assume a generic field $X$, defined on a pixelized 2D map such as: $\tilde{X}(\hat{\textbf{n}}) = w^{\rm X}(\hat{\textbf{n}})X(\hat{\textbf{n}})$, where the weight function $w^{\rm X}(\hat{\textbf{n}})$ represents the effect of the mask. The pseudo angular power spectrum of two generic masked fields $\{X, Y\}$ is defined as
    \begin{equation}
        \tilde{C}_{\ell}^{\rm XY} = \frac{1}{2\ell+1}\sum_{m = -\ell}^{\ell} \Tilde{a}^{\rm X}_{\ell m}\Tilde{a}^{\rm Y*}_{\ell m},
    \end{equation}
    where $\Tilde{a}^{\rm X}_{\ell m}$ are the harmonic coefficients of map $X$. This estimator is biased due to the mode coupling induced by the presence of the mask, but it can be related to the true underlying power spectrum via:
    \begin{equation}
    \label{eq:pcls}
        \langle \tilde{C}_{\ell}^{\rm XY} \rangle = \sum_{\ell'} M_{\ell \ell^{\prime}}^{\rm XY} C^{\rm XY}_{\ell^{\prime}} + \tilde{N}_{\ell}^{\rm XY}.
    \end{equation}
    The term $\tilde{N}_{\ell}^{XY}$ is the noise pseudo angular power spectrum and it is different from zero for $X=Y$.
    In Equation~\ref{eq:pcls}, $M_{\ell \ell^{\prime}}^{XY}$ is the mode-coupling matrix, which depends exclusively on the pseudo cross-spectrum of the masks $\tilde{W}_{\ell}^{XY}$:
    \begin{equation}
       M_{\ell \ell^{\prime}}^{\rm XY} = \frac{(2\ell^{\prime}+1)}{4\pi} \sum_{\ell^{\prime\prime}}(2\ell^{\prime\prime}+1) \tilde{W}_{\ell''}^{\rm XY}
       \wtj{\ell}{\ell'}{\ell''}{0}{0}{0}^2.
    \end{equation}
    The unbiased power spectrum estimator is then:
    \begin{equation}
    \label{eq:true_cls}
        \hat{C}_{\ell}^{\rm XY} = \sum_{\ell^\prime} (M^{\rm XY})^{-1}_{\ell \ell^{\prime}} [\tilde{C}^{\rm XY}_{\ell^{\prime}} - \tilde{N}_{\ell^\prime}^{\rm XY}].
    \end{equation}
    To further account for the loss of information induced by the masked sky, the pseudo-$C_{\ell}$ is divided into band powers. In our analysis, we binned all the cross- and auto-spectra using the scheme proposed by \cite{GG_2021}: equally spaced bins with $\Delta \ell= 30$ in the range $\ell= [2,240]$ and logarithmic bins for greater $\ell$s, with $\Delta \log_{10}\ell = 0.055$.
    In our analysis, we used the {\tt NaMaster} code for the computation of power spectra and the analytical derivation of their covariance matrices, following the implementation described in \citep{GarciaGarcia_2019, Garcia-Garcia_2019, Nicola_2021}. The covariance matrix is computed by considering each field (i.e. $\kappa$ and $g$ in our analysis) to be a Gaussian random field and correcting for the mode-coupling induced by the presence of the cut sky. This correction is performed using the narrow-kernel approximation (NKA), where the width of the mode-coupling kernel is much smaller than the range of multipoles over which power spectra vary significantly.
    
    The mode-coupled noise bias for the quasar auto-correlation $\tilde{N}^{gg}_{\ell}$, caused by Poisson shot-noise, can be computed analytically as: 
    \begin{equation}
        \tilde{N}^{gg}_{\ell} = \frac{\langle w\rangle_p \Omega_p}{\bar{N}}.
    \end{equation}
    In the previous equation, the mask (i.e. the selection function in this analysis) is averaged over the sky, $\Omega_p$ is the solid angle of each {\tt HEALPix} pixel and $\bar{N}$ is the mean number of quasars per pixel.
    
    We applied scale cuts to both cross- and auto-spectra. The minimum value of multipole, $\ell_{\rm min} = 30$, is set to exclude the range of multipoles in which the galaxy auto-correlation may be dominated by systematics due to dust, stellar contamination and the \gaia scanning pattern, as described in \cite{Alonso_2023}. This cut effectively removes the first band power in the power spectrum, and was selected in \cite{Alonso_2023} by observing that it was the only $\ell$-bin that changed significantly after linearly deprojecting the contaminants listed above at the pixel level. While this demonstrates that the selection function probably accounts for systematic contamination on smaller scales, it is worth noting that it may be possible to extract information from angular scales larger than that associated with $\ell=30$ by, e.g., estimating the power spectrum on narrower bandpowers below that $\ell$. This was not necessary in this analysis, as most of the information is concentrated on smaller angular scales, but other science cases in which large-scale information is critical (e.g. primordial non-Gaussianity) may benefit from doing so. The maximum multipole used in the analysis $\ell_{\rm max}$ (i.e. the smallest angular scale included) is given, for each redshift bin, by $\lmax = k_{\rm max}\chi(\bar{z})$, where $k_{\rm max} = 0.15$ Mpc$^{-1}$, and the comoving distance $\chi$ is computed at the mean redshift of each bin. The resulting values of $\lmax$ for the three redshift bins used in this work are $\lmax =[388, 709, 950]$. These scale cuts leave a total of 11, 16, and 18 bandpowers in the power spectra involving the first, second, and third redshift bins, respectively.
    
  \subsection{Likelihood}
  \label{sec:methods.likelihood}
    \begin{table}
        \centering
        \begin{tabular}{|c|c|}
        \hline
            Parameter & Prior \\
            \hline
            $\sigma_8$ & {\it U}(0.50, 1.20) \\
            $\Omega_M$ & {\it U}(0.05, 0.70) \\
            $h$ & {\it U}(0.40, 1.00) \\
            $b_g^i$ & {\it U}(0.1, 3.00) \\
            $\Delta(\bar{z}_i)$ & {\it U}(-1.00, 1.00)\\
            \hline
        \end{tabular}
        \caption{Prior distribution for the parameters used for the different analyses carried out in this work. $U(a,b)$ represents a uniform distribution.}
        \label{tab:priors}
    \end{table}
    To constrain the free parameters of our model from the measured cross- and auto-spectra, we use a Gaussian likelihood:
    \begin{equation}
    \label{eq:likelihood}
        -2\log p({\bf d} | \pmb{\theta})  \equiv \chi^2 + K = ({\bf t}(\pmb{\theta}) - {\bf d})^T {\sf Cov}^{-1}({\bf t}(\pmb{\theta}) - {\bf d}) + K,
    \end{equation}
    where $K$ is an arbitrary constant, ${\bf d}$ is the data vector storing all the estimated $C_{\ell}$s, ${\bf t}(\pmb{\theta})$ is the theoretical prediction for ${\bf d}$ given a set of parameters ${\pmb \theta}$ and ${\sf Cov}^{-1}$ is the inverse of the covariance matrix described in Section~\ref{ssec:methods.power_spectra}. Concretely, in our fiducial analysis, the data vector contains the galaxy autocorrelation and cross-correlation with $\kappa$ for the three redshift bins described in Section~\ref{ssec:methods.power_spectra} (i.e. a total of six power spectra), with the scale cuts described in Section~\ref{ssec:methods.power_spectra}. The length of ${\bf d}$ is therefore $N_d=72$ in our fiducial analysis. Note, however, that at times we will study only subsets of this data vector (e.g. only cross-correlations), or different numbers of redshift bins (e.g. see Section~\ref{ssec:res.fg}).
    
    In all the cases studied here, the free parameters of our model will, at least, contain the bias parameters $b_g^i$ for all the redshift bins that are being analysed. In addition to these, we will also consider variations in different sets of cosmological parameters, depending on the model being studied.
    The first two models are labelled respectively {\tt CosmoFix} and {\tt CosmoMarg}. The {\tt CosmoFix} model is defined by fixing the six $\Lambda$CDM cosmological parameters to the best-fit values found by \planck \cite{Planck_2020b}. The {\tt CosmoMarg} model has three free $\Lambda$CDM parameters, $\{\sigma_8,\,\Omega_M,\,h\}$, fixing the physical baryon density $\Omega_bh^2$, and the scalar spectral index $n_s$ to the \planck best-fit values while, as mentioned in Section~\ref{ssec:methods.growth_evolution}, we set the mass of neutrinos to zero. 
    In the third model explored here, labelled {\tt CosmoGrowth}, we aim to reconstruct deviations from $\Lambda$CDM in the growth history. As described in Section~\ref{ssec:methods.growth_evolution}, we quantify this with three parameters, $\Delta(\bar{z}_i)$, which quantify the relative deviation with respect to the growth factor predicted by the \planck best-fit $\Lambda$CDM model at the mean redshifts of the three redshift bins considered here. Since we do not enforce any normalisation on the linear growth factor in this case, the value of $\sigma_8$ is absorbed by these parameters (in other words, we fix the value of $\sigma_8$ used in the template linear power spectrum of Equation~\ref{eq:lingrowth}). In addition to $b_g^i$, the free parameters of the {\tt CosmoGrowth} model are therefore $\{\Omega_M,h,\Delta(\bar{z}_1),\Delta(\bar{z}_2),\Delta(\bar{z}_3)\}$. The different priors used in our analyses are listed in Table~\ref{tab:priors} and, in particular, the priors on $\Delta(\bar{z}_i)$ were designed to ensure that $D(z)$ remains non-negative throughout. As in \cite{Alonso_2023}, we include a BAO prior, using data from BOSS and eBOSS \cite{BOSS:2016wmc,eBOSS_BAO}.
    
    To explore the multidimensional likelihood, we use a Monte-Carlo Markov Chain algorithm as implemented in {\tt Cobaya}\footnote{\url{https://cobaya.readthedocs.io/en/latest/}} \cite{Torrado_2021}, with a convergence criterion of $R-1<0.01$, where $R$ is the Gelman-Rubin parameter \citep{GelmanRubin}. \newv{In what follows, when referring to ``best-fit'' parameters from our data, these will correspond to the MCMC sample with the highest log-posterior.} All theoretical calculations were carried out using the Core Cosmology Library (CCL\footnote{\url{https://github.com/LSSTDESC/CCL}},  \cite{Chisari_18}).
     
\section{Results}\label{sec:results}
  \subsection{Measured power spectra}
\begin{figure}[!h]
        \centering
        \includegraphics[width = 0.75\textwidth]{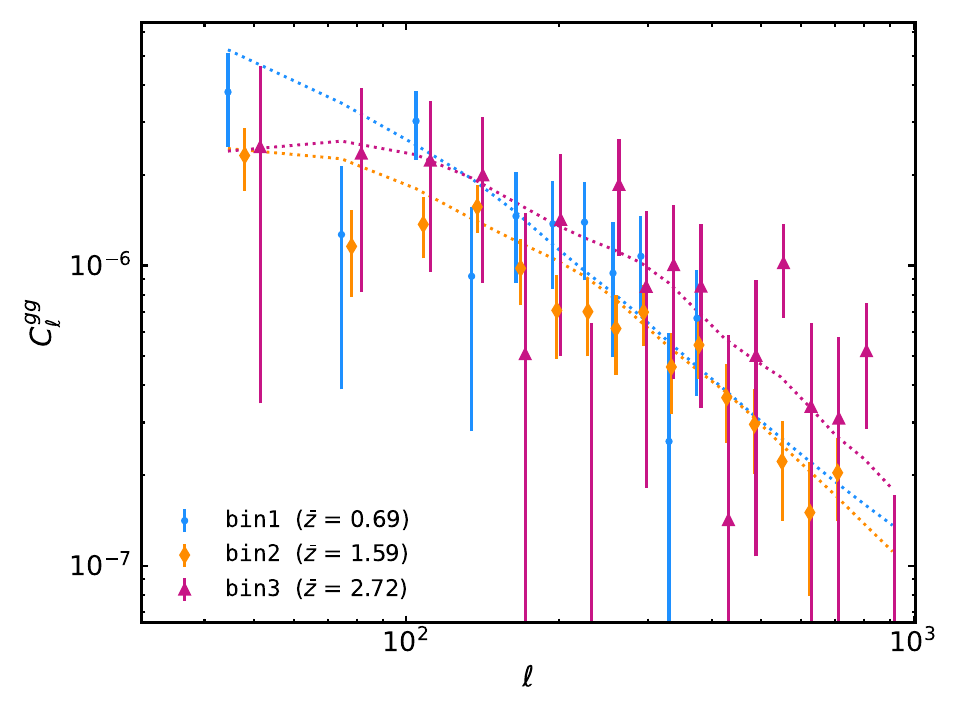}
        \caption{Auto angular power spectrum of the quasars for three \quaia bins, as indicated in the legend, in the range of multipoles from $\ell>30$. For each redshift bin, we applied the $\ell_{\rm max}$ cut defined in Section~\ref{ssec:methods.power_spectra}. Dashed lines represent the theoretical predictions obtained, for each bin, by considering the best-fit bias parameters for the {\sf GMV} case. The estimated power spectra are plotted
        with an offset along the $x$-axis for a better visualization.}
        \label{fig:cls_gg}
    \end{figure}

\begin{figure}
        \centering
        \includegraphics[width =\textwidth]{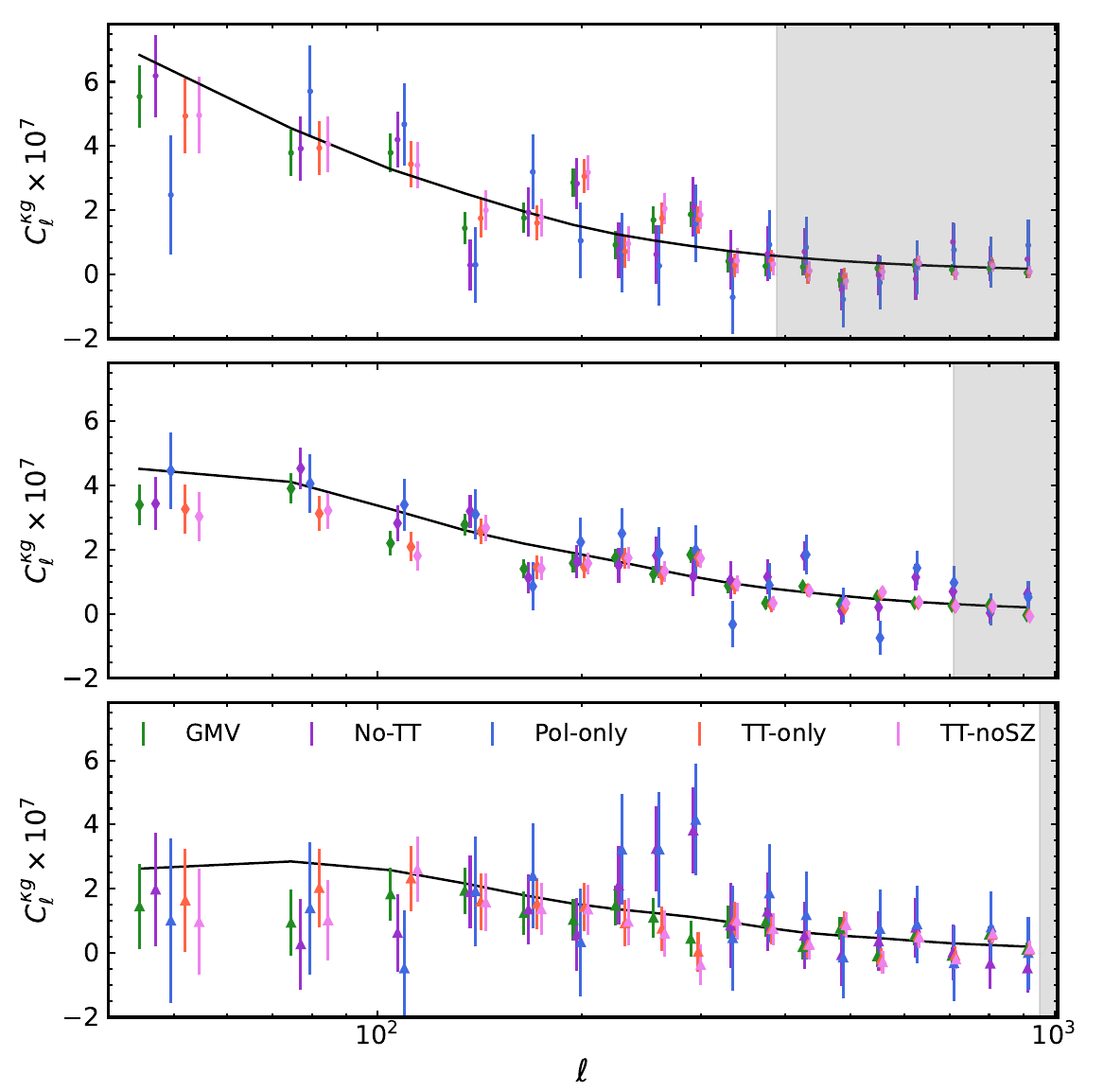}
        \captionof{figure}{\quaia - CMB lensing cross angular power spectrum for different redshift bins ({\tt bin1}: top panel, {\tt bin2}: central panel and {\tt bin3}: bottom panel). In each plot, different colours show the cross-correlation with the various CMB lensing reconstructions considered in the analysis while solid black lines represent the theoretical predictions. The estimated power spectra are plotted with an offset along the $x$-axis for a better visualisation.}
        \label{fig:cls_kg}
    \end{figure}

    We measure the quasar auto-correlations in the three redshift bins described in Section~\ref{ssec:data.quaia} and their cross-correlation with the different $\kappa$ maps described in Section~\ref{ssec:data.cmb}. The resulting auto-spectra are shown in Figure~\ref{fig:cls_gg} in blue, orange and magenta for the first, second, and third bins, respectively, while their cross-spectra with five of the lensing reconstruction maps used here are shown in Figure~\ref{fig:cls_kg}. In both power spectra, the statistical uncertainties in these measurements are significantly larger for bin 3 than for bin 2. This is due to the contribution from shot noise to the error and is caused by the lower number density of sources in the highest redshift bin.

   Nevertheless, all power spectra are detected with a high signal-to-noise ratio in all bins. The detection significance is estimated as $\sqrt{\chi^2_{\rm null} -\chi^2_{\rm bf}}$, where $\chi^2_{\rm null}$ is the value of $\chi^2$ (see Equation~\ref{eq:likelihood}) for a null prediction (${\bf t}= {\bf 0}$), and $\chi^2_{\rm bf}$ is the best-fit value in the {\tt CosmoMarg} model (see Section~\ref{sec:methods.likelihood}). The detection significances of each individual power spectrum are summarised in Table~\ref{tab:snr}. In most cases, we find that the cross-spectrum has a larger statistical significance than the auto-spectrum, since the latter is more strongly affected by shot noise. The most sensitive cross-correlation measurements correspond, unsurprisingly, to the {\sf GMV} $\kappa$ map. Since the sensitivity of the \planck CMB lensing data is dominated by temperature information, the  {\sf TT-only} and  {\sf TT-noSZ} cross-correlations achieve better sensitivity than the  {\sf Pol-only} and {\sf No-TT} maps. Nevertheless, the {\sf No-TT} map, which recovers some of the temperature information, while remaining largely immune to unpolarized foregrounds, is significantly more sensitive than the {\sf Pol-only} map employed in \cite{Alonso_2023} to quantify a potential foreground contribution to the \quaia cross-correlation at high redshifts. Thus, our fiducial results will largely focus on the {\sf GMV} and {\sf No-TT} maps, as they represent the most sensitive CMB lensing reconstructions, achieving the best compromise between sensitivity and robustness to potential extragalactic foregrounds, respectively. However, we will also discuss the results obtained with the other CMB lensing reconstruction methods in Section \ref{ssec:res.fg}

    As hinted at in Section \ref{ssec:data.cmb}, and in agreement with the findings of \cite{Alonso_2023}, we can see that, particularly in the case of bin 2, the amplitude of the cross-correlation with the {\sf No-TT} and {\sf Pol-only} $\kappa$ maps is consistently higher than all other $\kappa$ maps. We will explore this effect further in Section~\ref{ssec:res.fg}. \newv{Appendix \ref{app:lcdm} presents the constraints on standard $\Lambda$CDM parameters found from these measurements, and quantifies the differences with respect to the results presented in \cite{Alonso_2023} due to changes in the \quaia catalogue and selection function, the choice of redshift binning, and the impact of potential Galactic and extragalactic contamination (see Section \ref{ssec:res.fg}).}
    \begin{table}
      \centering
      \begin{tabular}{|c|c|c|c|c|c|c|}
        \hline
        & $C_{\ell}^{gg}$ &$C_{\ell}^{\kappa g, \rm {\sf No-TT}}$  & $C_{\ell}^{\kappa g, \rm {\sf GMV}}$ & $C_{\ell}^{\kappa g, \rm  {\sf Pol-only}}$ & $C_{\ell}^{\kappa g, \rm  {\sf TT-only}}$ & $C_{\ell}^{\kappa g, \sf no-SZ}$\\
        \hline
        {\tt bin1} & 7.5 & 8.7 & {\bf 13.5} & 5.4 & 11.1 & 11.4 \\
        \hline
        {\tt bin2} & 14.7 & 13.5 & {\bf 20.0} & 10.0 & 15.5 & 15.1 \\
        \hline
        {\tt bin3} & 5.9 & 5.1 & {\bf 5.7} & 2.9 & 5.0 & 3.8 \\
        \hline
        All bins & 17.3 & 16.2 & {\bf 24.7} & 11.6 & 19.8 & 19.4 \\
        \hline
      \end{tabular}
      \caption{Significance of the detection $S/N\equiv\sqrt{\chi^2_{\rm null} -\chi^2_{\rm bf}}$ in unit of equivalent $\sigma$ for a Gaussian distribution for the different bins. Second column: results for the auto-spectrum. Third to seventh column: results for the cross-power spectra between \quaia bins and the lensing reconstructions used in the analysis. Values in bold indicate the highest significance obtained with the different CMB lensing reconstruction.}\label{tab:snr}
    \end{table}

  \subsection{Growth reconstruction}\label{ssec:res.growth}
\begin{figure}[h!]
      \centering
      \includegraphics[width = 0.85\textwidth]{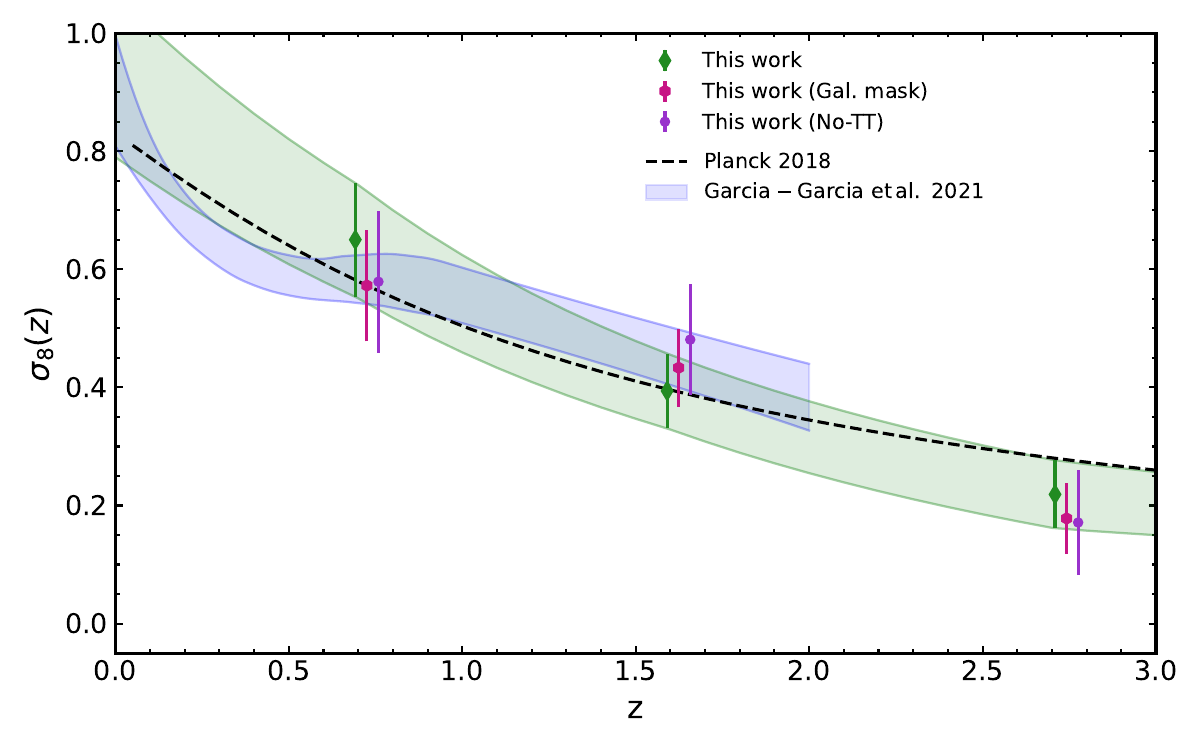}
      \caption{Amplitude of the matter fluctuations $\sigma_8$ as a function of redshift. The dashed line is obtained with the fiducial Planck cosmological parameters, while the navy shadowed area represents the constraints found by \cite{GarciaGarcia_2019}. The points in the plot represent the measurements obtained with the Gaussian likelihood for three different CMB lensing reconstructions: {\sf GMV} (green diamonds), {\sf GMV Gal. mask} (magenta hexagons) and {\sf No-TT} (purple dots).}
      \label{fig:s8_z}
    \end{figure}

    We use the power spectrum measurements presented in the previous section, to reconstruct the growth history following the procedure described in Section~\ref{ssec:methods.growth_evolution}. As described there, we constrain the deviations with respect to the fiducial $\Lambda$CDM growth history $\Delta(\bar{z}_i)$ and, from them, place constraints on $\sigma_8(z)$. The results are shown in Figure~\ref{fig:s8_z}, where the green band shows the $1\sigma$ constraints found for the {\sf GMV} CMB lens reconstruction, while the magenta hexagons and purple circles show results for the {\sf GMV-Gal. mask}, and {\sf No-TT} cases, respectively.
    The dashed line shows the constraints of \planck (\cite{Planck_2020b}), while the navy band shows the constraints found by \cite{GarciaGarcia_2019} by combining a larger suite of galaxy clustering, cosmic shear, and CMB lensing datasets, covering mostly lower redshifts than those explored here.
    With \quaia, we are able to extend the constraints of \cite{Alonso_2023} to higher redshifts, adding a new data point at $z\sim2.7$. We find that, throughout the entire redshift range covered by \quaia ($z\lesssim3$), our constraints are in reasonably good agreement with the \planck best-fit model. 
    
\begin{figure}[h]
        \centering
        \includegraphics[width = 0.80\textwidth]{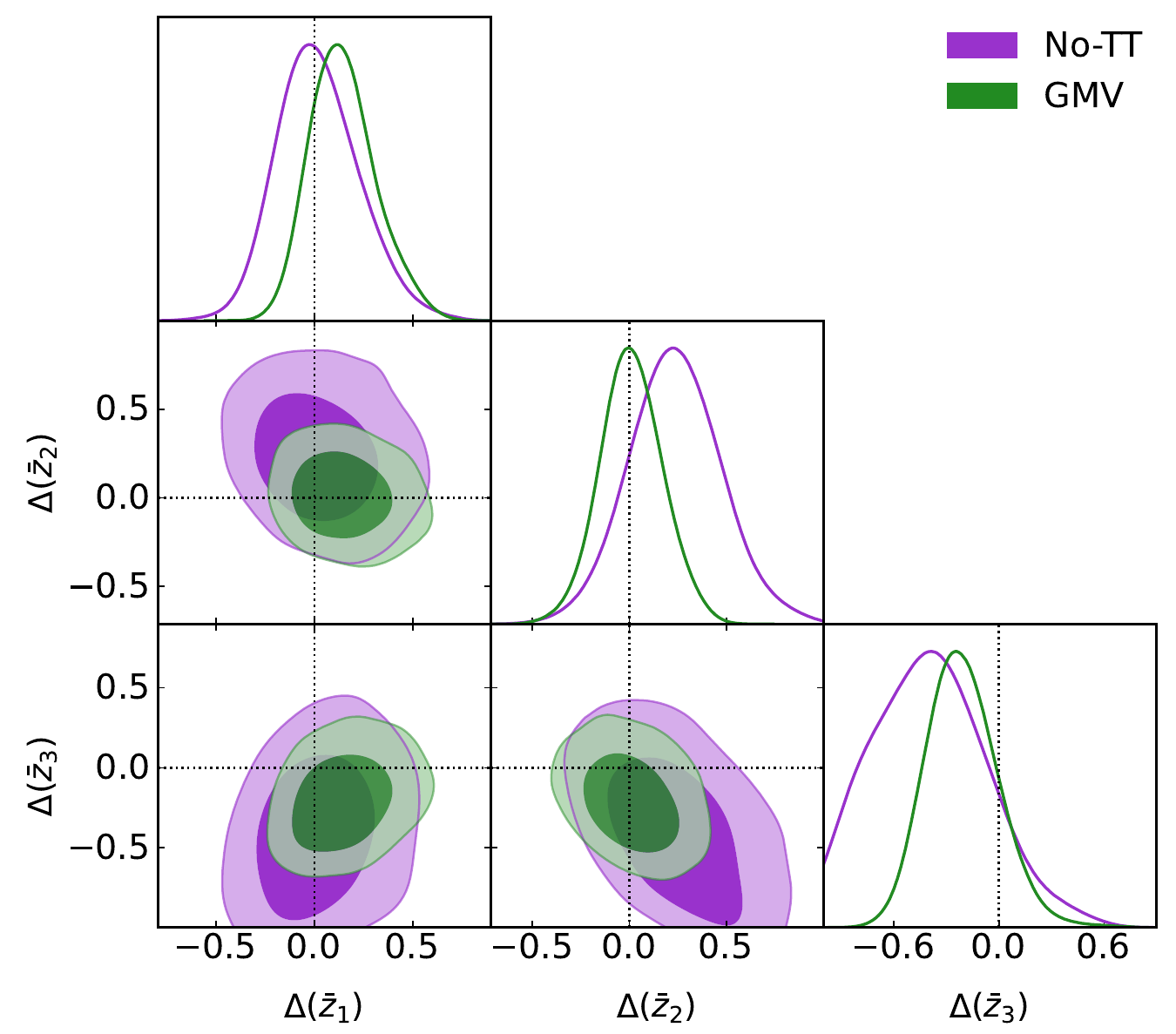}
        \caption{Marginalized posterior distribution for $\Delta_i$ parameters obtained from the {\sf GMV} (green) and {\sf No-TT} (purple) CMB lensing maps. Results are obtained within the {\tt CosmoGrowth} model. Dashed lines represent the value of $\Delta(\bar{z}_i) = 0$ predicted by $\Lambda$CDM cosmology.}
        \label{fig:delta_i}
    \end{figure}

    The multi-dimensional constraints on the three $\Delta(\bar{z}_i)$ parameters recovered by this analysis are shown in Figure~\ref{fig:delta_i}. 
    The marginalized constraints on these parameters for the three redshift bins are: 
    \begin{equation*}
    \begin{split}
      &\Delta(\bar{z}_{i})_{\sf GMV} = \{0.15 \pm 0.17,\hspace{6pt}0.01 \pm 0.16,\hspace{6pt} -0.21 \pm 0.21\}, \\
      &\Delta(\bar{z}_{i})_{\sf GMV, Gal. mask}= \{0.01 \pm 0.17,\hspace{6pt} 0.11 \pm 0.17,\hspace{6pt} -0.36 \pm 0.22\}\\
      &\Delta(\bar{z}_{i})_{\sf No-TT}= \{0.02 \pm 0.21,\hspace{6pt} 0.23 \pm 0.23,\hspace{6pt} -0.38 \pm 0.32\}.
      \end{split}
    \end{equation*}
    all of which are compatible with zero at less than $\sim1.5\sigma$. The associated values of $\sigma_8(z)$ are
    \begin{equation}\label{eq:s8z}
    \begin{split}
        &\sigma_8(\bar{z}_i)_{\sf GMV}=\{0.65\pm0.10,\hspace{6pt}0.39\pm0.06,\hspace{6pt}0.22\pm 0.06\}\\
        &\sigma_8(\bar{z}_i)_{\sf GMV, Gal. mask}=\{0.57\pm0.09,\hspace{6pt}0.43\pm0.07,\hspace{6pt}0.18\pm0.06\}\\
        &\sigma_8(\bar{z}_i)_{\sf No-TT}=\{0.57\pm0.12,\hspace{6pt}0.48\pm0.09,\hspace{6pt}0.17\pm0.09\}.
      \end{split}
    \end{equation}
    As shown in Figure~\ref{fig:delta_i}, particularly in the case of the {\sf GMV} map, the constraints between different $\Delta(\bar{z}_i)$s are largely uncorrelated, and the posterior distribution is rather Gaussian. The marginalized errors, quoted above, are therefore representative of the uncertainties on the growth factor at different redshifts measured by \quaia. 
    
    The agreement with the growth history predicted by \planck can be further quantified by fitting the recovered values of $\sigma_8(z)$ at the three redshift nodes (i.e. at $\bar{z}_i$) with the prediction of \planck with a free amplitude parameter $A_{\sigma_8}$, assuming that the measured values of $\Delta(\bar{z}_i)$ are Gaussian distributed, with a covariance matrix recovered from the MCMC chains. The resulting values of $A_{\sigma_8}$ are
    \begin{equation*}
      A_{\sigma_8}^{\sf GMV} = 0.97 \pm 0.09, \quad
      A_{\sigma_8}^{\sf GMV, Gal. mask} = 0.96 \pm 0.09, \quad
      A_{\sigma_8}^{\sf No-TT} = 1.12 \pm 0.09
    \end{equation*}
    compatible with $A_{\sigma_8}=1$ within $\sim1\sigma$. More in detail, the upward shift in $\sigma_8$ when discarding temperature auto-correlations in the CMB lensing reconstruction, which was described in the previous Section, is only evident here in the intermediate redshift bin ($z\sim1.6$), in which the {\sf No-TT} analysis recovers a value of $\Delta(z)$ that is $\sim1\sigma$ higher than the {\sf GMV} value\footnote{As described in Section \ref{ssec:methods.growth_evolution}, $D(z)$ and $\sigma_8(z)$ are related via $\sigma_8(z)=\sigma_8^{\rm fid}D_{\rm fid}(z)[1+\Delta(z)]$.}. This weak evidence disappears at both lower and higher redshifts. In fact, the new measurement at $z\sim2.7$ is $\sim1\sigma$ low compared to \planck for the {\sf GMV} map, and $\sim1.3\sigma$ low for the {\sf No-TT} map. \newv{Using the more restrictive Galactic mask does not lead to significant changes in these results.} We explore this in more detail in Section~\ref{ssec:res.fg}.

    Overall, we find that the growth of structure over the redshift range covered by \quaia, which spans the last $\sim11.8$ billion years of cosmic history, is well described by the best-fit $\Lambda$CDM model found by \planck.

  \subsection{Bias evolution}\label{ssec:res.bias_evol}
\begin{figure}[h]
        \centering
        \includegraphics[width = 0.75\linewidth]{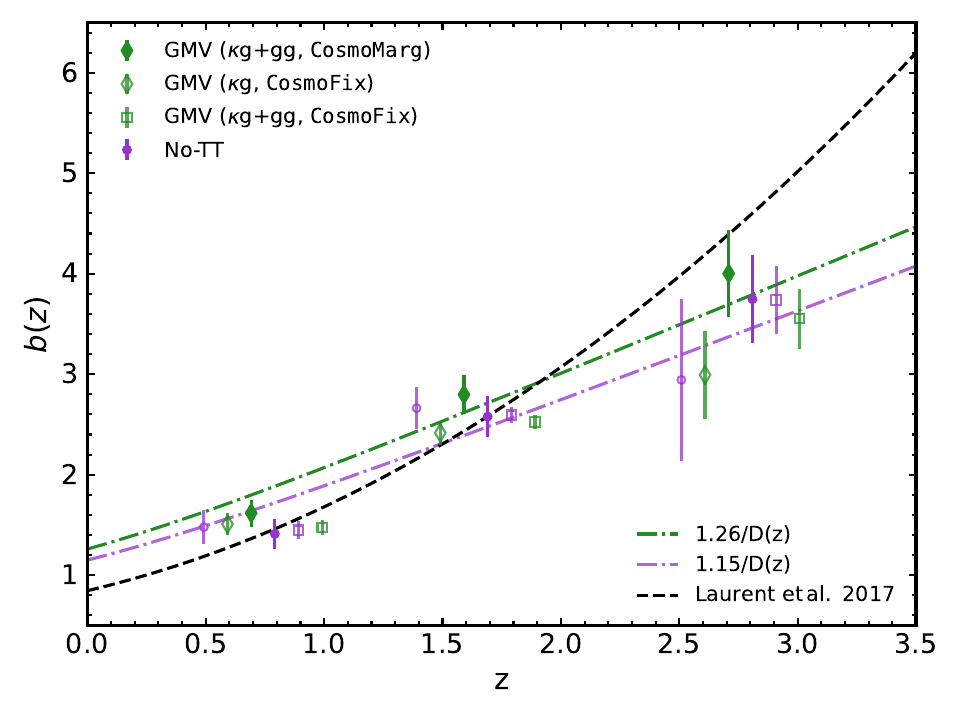}
        \caption{Quasar bias evolution as a function of redshift. Different colours show results obtained with different CMB lensing maps ({\sf GMV} in green, {\sf No-TT} in purple). Filled markers represent results obtained with the quasars' auto- and cross-correlation with CMB lensing, while empty diamonds and circles are obtained with the CMB lensing cross-correlation alone. Lastly, empty square markers are obtained using the full data vector, namely $\kappa g+gg$ but within the {\tt CosmoFix} cosmological model.  The bias is evaluated at the mean redshift of each bin. The dashed black line represents the fiducial eBOSS bias model of Equation~\ref{eq:bias} for which $b_g =1$, while the coloured dashed lines represent the bias model proposed here $b(z) = b_0/D_{\rm fid}(z)$, where $b_0$ is the best-fit for the {\tt CosmoMarg} case obtained considering $\kappa g+gg$. These bias models display a milder evolution at high redshifts (see the text for more details). }
        \label{fig:bi_zi}
    \end{figure}

    Considering the fiducial $N(z)$ for the \quaia bins to be the true description of the quasar redshift distribution, we can constrain the bias evolution of the sources by extracting information from the power spectra. We use the likelihood defined in Section~\ref{sec:methods} and the bias is evaluated in each redshift bin using Equation~\ref{eq:bias}. In particular, we perform two different tests, depending on the data vector and the cosmological model considered. In the first one, the data vector $\boldsymbol{d}$ of Equation~\ref{eq:likelihood} is given by the combination of the cross- and auto- power spectra of the \quaia bins, and we assumed a {\tt CosmoMarg} model. The results are shown with filled markers in Figure~\ref{fig:bi_zi} for the {\sf GMV} and {\sf No-TT} cases. 
    The evolution of the bias is generally in qualitative agreement with the fiducial eBOSS-based model (dashed black line), although some differences must be noted. First, the values recovered at low redshifts are consistently higher than those in the eBOSS model. The fact that the \quaia bias is potentially higher than that of the eBOSS model is not entirely surprising, since the \quaia sample is slightly brighter than the eBOSS sample. Perhaps more interestingly, we find that the quasar bias grows more slowly than the eBOSS model at higher redshifts. This, again, is likely due to differences between the eBOSS and \quaia samples. 
    
    Given the previous discussion, it is interesting to consider the predicted value and evolution of the quasar bias under the assumption that the best-fit cosmological model found by \planck is the true underlying cosmology. In particular, fixing cosmology allows us to measure the quasar bias using only the quasar-CMB lensing cross-correlation, which is potentially less sensitive to systematic contamination in the quasar overdensity maps, as well as to uncertainties in the redshift distribution of the sample (which may be significant at high redshifts) \citep{Alonso_2021}. \newv{The outcome of this analysis is shown with hollow markers in Figure~\ref{fig:bi_zi}, squares for the combination of $C_\ell^{gg}$ and $C_\ell^{\kappa g}$, and circles for $C_\ell^{\kappa g}$ alone. We find that the main effect of fixing the cosmological model is an overall downward shift in the recovered $b(z)$ by about 1 to $1.5\sigma$. This makes sense, since the value of $\sigma_8$ recovered by the \quaia data is $\sim1\sigma$ lower than the \planck best fit (see \cite{Alonso_2023} and Appendix \ref{app:lcdm}), which must be compensated for by lowering the amplitude of $b(z)$. It is worth noting that this effect is present for both $C_\ell^{gg}+C_\ell^{\kappa g}$ and for $C_\ell^{\kappa g}$ alone, showing that the amplitudes of both power spectra are compatible within the preferred $\Lambda$CDM model. In this case, we also recover a slower evolution for $b(z)$ at high redshifts when compared with the eBOSS model. The results found using the {\sf No-TT} $\kappa$ map (shown in purple) are generally in good agreement with our fiducial constraints, although with a $\sim0.5-1\sigma$ downwards shift, compatible with the higher value of $\sigma_8$ recovered by this map. Thus, we conclude that this result is not strongly driven by contamination from extragalactic foregrounds in the CMB lensing map. Repeating our analysis imposing the 40\% Galactic mask (not shown in the figure) recovers results that are in good agreement with the fiducial case, reassuring us that our results are not significantly affected by Galactic systematics (neither in \quaia nor in the $\kappa$ maps).}
    
    \newv{Finally, we provide a model for the bias of the \quaia sample based on a simpler parametrization of the form $b(z)=b_0/D_{\rm fid}(z)$, where $D_{\rm fid}(z)$ is the linear growth factor for the best-fit \planck cosmology, and $b_0$ is a free parameter. This may be convenient for future studies using this sample. For the combination of $gg$ and $\kappa g$ with free cosmological parameters, we find:}
    \begin{equation}
      b_0^{\sf GMV}=1.26\pm  0.08,\hspace{12pt}b_0^{\sf No-TT}=1.15\pm0.09.
    \end{equation}
    \newv{The resulting best-fit models are shown in Figure~\ref{fig:bi_zi} with dotted and dashed lines.}
    \newv{Using only the galaxy-lensing cross-correlation, and fixing all cosmological parameters to the \planck best fit, we find}
    \begin{equation}
      b_0^{GMV} = 1.12\pm  0.04,\hspace{12pt}b_0^{\sf No-TT} = 1.17\pm0.08.
    \end{equation}

  \subsection{Foreground systematics in CMB lensing reconstruction}
  \label{ssec:res.fg}
   We now perform an analysis which aims to determine the possible presence of the foreground contamination in the \planck CMB lensing maps, as well as its potential origin. Ref.~\cite{Alonso_2023} reported a potentially significant difference in the amplitude of the \quaia-CMB lensing cross-correlation using different $\kappa$ maps at high redshifts, which we also saw hints of in the previous sections, and in Appendix \ref{app:lcdm}. Isolating the range of redshifts over which this contamination is more relevant would be interesting in order to obtain a more significant detection of it, and to establish its potential extragalactic origin. For this reason, in this case we divide the total redshift distribution of the \quaia sources into six redshift bins, ensuring that each bin contains a suitable number of sources to mitigate excessive shot noise. Specifically, the redshift ranges used to define these bins are: $z_{\rm edges} = [0.0, 0.5, 1.0, 1.5, 2.0, 2.5, 5]$. For each redshift bin, we also computed the associated selection function $w$ to take into account the new redshift ranges.

    In this case, we did not perform a full likelihood analysis for each map, but instead we simply used the galaxy-$\kappa$ cross-correlation with fixed cosmological parameters to measure the galaxy bias as a way to quantify the amplitude of this cross-correlation. Since, in this case, the model is linear in $b_g$, it can be computed analytically as:
    \begin{equation}\label{eq:bg_kg}
      b_g = \frac{{\bf t}^T{\sf Cov}^{-1}{\bf d}}{{\bf t}^T{\sf Cov}^{-1}{\bf t}},\hspace{12pt} \sigma(b_g) = \frac{1}{\sqrt{{\bf t}^T{\sf Cov}^{-1}{\bf t}}},
    \end{equation}
    where ${\bf t}$ is the theoretical cross-power spectrum calculated for $b_g=1$, ${\bf d}$ is the measured cross-spectrum, and ${\sf Cov}$ is the associated covariance matrix.

\begin{figure}
        \centering
        \includegraphics[width = \linewidth]{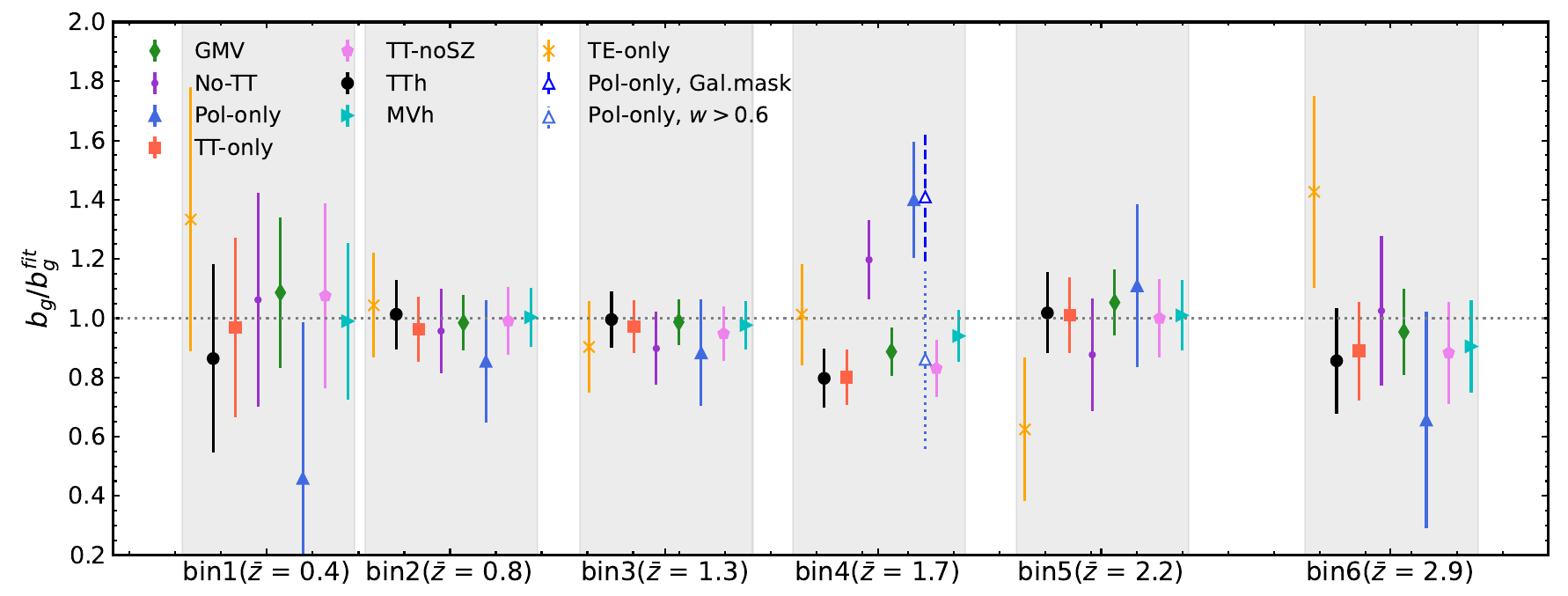}
        \caption{Relative amplitude of the CMB lensing cross-correlation as function of mean redshift. Colored markers represent constraints from different CMB lensing reconstructions, as indicated in the legend. Each cloud of points is centered on the mean redshift value of the redshift bin, as indicated by the x-axis label. In the fourth redshift bin, centered at $z=1.7$, we also display results for the {\sf Pol-only} map imposing the 40\% Galactic mask (dashed error bar), and a more conservative selection function with $w>0.6$ (dotted error bar) (see text for more details).}
        \label{fig:bi_zi_6bins}
    \end{figure}

   The results for all eight different CMB $\kappa$ maps are shown in Figure~\ref{fig:bi_zi_6bins}. 
   To aid in visualizing these results, we present our measurements with a normalization factor determined as follows: we consider the {\sf No-TT} measurements (purple points in Figure~\ref{fig:bi_zi_6bins}) and fit a second-order polynomial function in redshift to them. Subsequently, we use the best-fit $b(z)$ polynomial obtained as the normalization factor for all the $b_g$ measurements.
   From this figure, it is visually evident that different CMB $\kappa$ maps are largely consistent with one another throughout the whole redshift range except in the fourth bin, centered around $z\simeq 1.7$. As in the case of the constraints on $\sigma_8$ found by \cite{Alonso_2023} (see also Appendix \ref{app:lcdm}), we find that the CMB lensing reconstructions involving $TT$ correlations recover a lower value of $b_g$ than {\sf Pol-only} and {\sf No-TT}, seemingly at $\sim2-3\sigma$ significance.

\begin{figure}
        \centering
        \includegraphics[width = 0.8\linewidth]{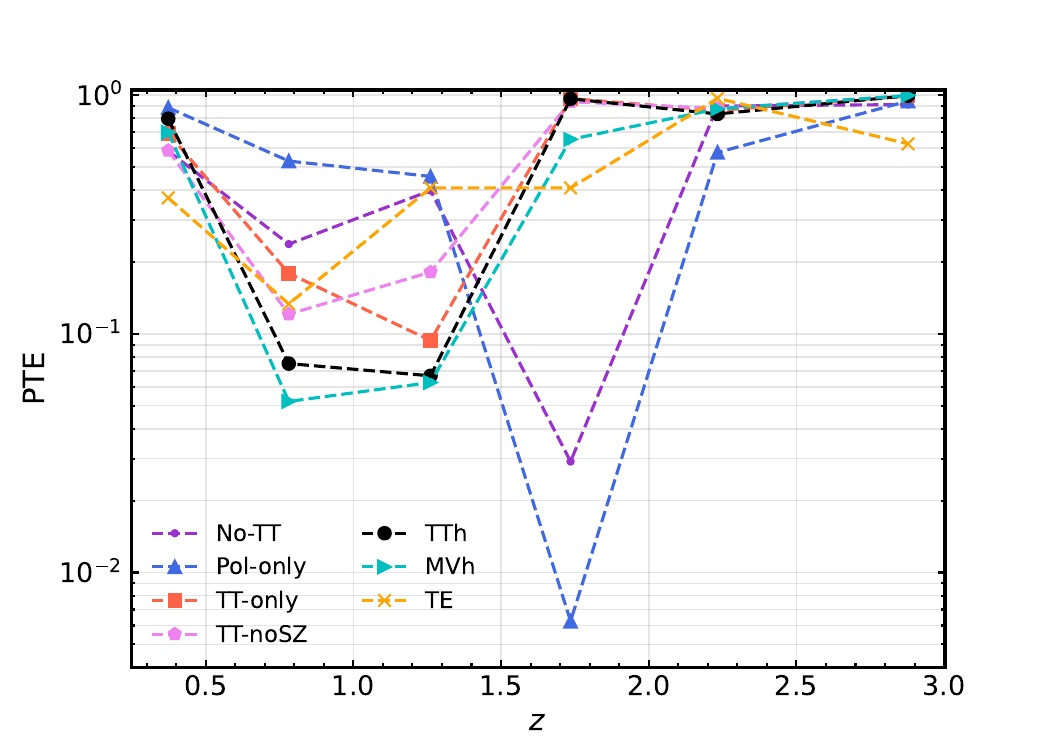}
        \caption{Probability to exceed of the difference between the amplitude of the $\kappa g$ cross-correlation found with the {\sf GMV} $\kappa$ map and all other lensing reconstruction maps considered in Figure \ref{fig:bi_zi_6bins}, as a function of redshift. }
        \label{fig:pte_gmv}
    \end{figure}

    The actual significance of these differences is hard to assess just from the statistical uncertainties shown in this figure, since the different CMB lensing maps were constructed from the same set of CMB maps, and thus the different estimates of $b_g$ are correlated. To quantify the statistical significance of these differences consistently, we made use of correlated simulations as follows. The \textit{Planck} PR4 lensing analysis delivered 480 independent realizations of simulated lensing reconstructions based on the signal and sky model of the \textit{Planck} FFP10\footnote{\url{https://wiki.cosmos.esa.int/planck-legacy-archive/index.php/HFI_sims}} simulations and the noise properties of the NPIPE maps. The 8 different types of  lensing reconstructions explored here are run on the same simulated data sets. As such, they share the underlying signal and noise in the CMB maps, while the resulting lensing reconstruction noise differs. For each of the 480 available realizations of noisy lensing reconstructions, we retrieved the corresponding input $\kappa$ signal realization. For each of the $\kappa$ maps, we generated a correlated Gaussian realization of the quasar overdensity map, $\delta_g$. To do so, we assumed the theoretical auto and cross-correlation with $\kappa$, the same cosmology as in the FFP10 simulations, as well as the fiducial bias model and redshift distribution for each bin. To simulate the noise contribution for the quasars, we added to $\delta_g$ a shot noise component modeled using random realizations from the \quaia catalogue matching the number of objects included in each redshift bin. We then computed the cross-spectrum $C_\ell^{\kappa g}$ for each realization and $\kappa$ reconstruction method, and estimated the value of $b_g$ as in Equation~\ref{eq:bg_kg}. Finally, we quantify the level of disagreement between different CMB lensing maps by calculating the fraction of simulated realizations for which a difference in the value of $b_g$ with respect to the one found for the corresponding {\sf GMV} map is larger than the value measured in the data. This corresponds to a simulation-based estimate of the consistency between the cross-correlations with different $\kappa$ maps in terms of a probability to exceed (PTE). When computing the PTE, we take into account the sign of the difference in the value of $b_g$ and, therefore, we considered the 2-sided probability. The probability-to-exceed values are shown in Figure \ref{fig:pte_gmv}. We find that the level of tension in the 4-th bin between the {\sf No-TT} map and the {\sf GMV} map, and between the {\sf Pol-only} map and the {\sf GMV} map, is relatively high, with probabilities of 0.03 and 0.006 respectively, corresponding to a 2-3$\sigma$ significance.
        
    Interestingly, this tension is not reproduced by the source-hardened maps ({\sf MVh} and {\sf TTh}) and, although the {\sf TE} map does recover a marginally larger value of $b_g$ (by about $0.5\sigma$ in the same direction as the {\sf Pol-only} and {\sf No-TT} maps), the level of disagreement with the {\sf GMV} map is much smaller.  Although this still does not rule out the CIB or other extragalactic foregrounds a potential cause of this disagreement (since their clustered component might be immune to source hardening), it is worth exploring Galactic foreground contamination in the polarization maps as an alternative possibility to explain this tension. To do so, \newv{we repeat our analysis using the 40\% Galactic mask from \planck (see Figure \ref{fig:footprints}). The resulting value of $b_g$ from the {\sf Pol-only} map, shown as a blue triangle with dashed error bars, is in good agreement with the result found with our fiducial mask. We performed a similar test using a more restrictive mask, defined by nulling pixels in which the selection function was $w\leq0.6$, and removing a substantially larger sky fraction. Although the value of $b_g$ recovered in this case (shown in Figure \ref{fig:bi_zi_6bins} as a triangle with dotted error bars) is in better agreement with the {\sf GMV} result, we find that this is mostly driven by statistical fluctuations at $\ell\gtrsim300$, where the cross-correlation is compatible with zero, whereas at low $\ell$, where the cross-correlation is actually detected, the measurement is in better agreement with the {\sf No-TT} and {\sf Pol-only} maps on the fiducial mask than with the {\sf GMV} map. Because of this, together with the significantly larger error bars of this measurement, due to the small sky fraction, we consider this to be compatible with a statistical fluctuation.}
    
    \newv{We thus find no conclusive evidence that the differences found between different redshift bins are caused by Galactic contamination in the CMB lensing maps. On the other hand, if extragalactic in origin, this potential systematic would evade source-hardened estimators, thus requiring it to have a significant clustered component. Although the CIB is an attractive candidate, it is not clear why its presence is not evident at higher redshifts, and our investigation is ultimately inconclusive.} Ultimately, repeating this analysis with CMB lensing maps constructed from other CMB experiments, such as the Atacama Cosmology Telescope (ACT, \cite{Qu_2023}), or the South Pole Telescope (SPT, \cite{Pan_2023}), will be vital to address this issue. We leave this analysis for future work.

\section{Conclusions}\label{sec:conclusions}
  In this work, we have studied the growth of cosmic structure as a function of time during the last 11.8 billion years, using the auto-correlation of the \quaia quasar sample in combination with its cross-correlation with CMB lensing from \planck. In addition to this, we have studied the redshift dependence of the linear bias of this sample, and quantified the potential presence of Galactic and extragalactic foreground contamination on the \planck CMB lensing maps.

\begin{figure}
    \centering
    \includegraphics[width = 0.8\linewidth]{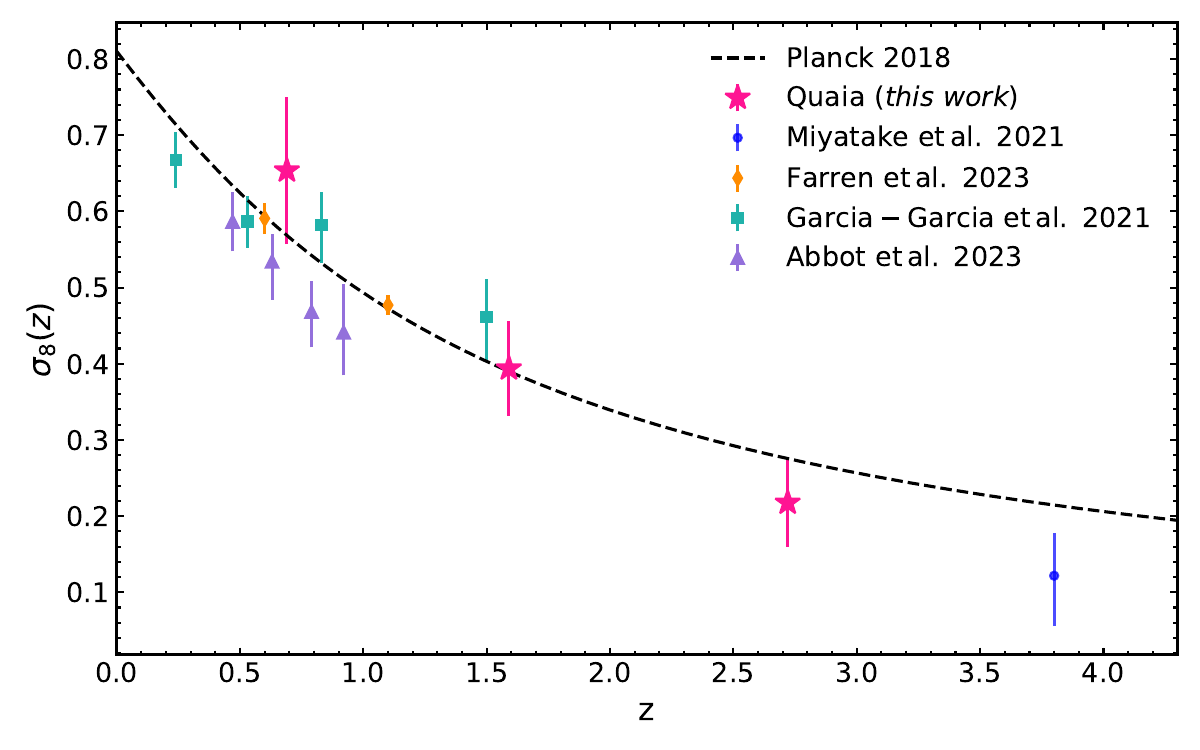}
    \caption{$\sigma_8(z)$ measurements as a function of redshift. We compare our results (magenta stars) with other measurements from previous studies (\citep{Miyatake_2022, Farren_2023, GarciaGarcia_2019, Abbot_2023}), as indicated in the legend. The dashed black line shows the $\Lambda$CDM prediction when using the \planck best-fit cosmological parameters.}
    \label{fig:s8_comp}
  \end{figure}

  Taking advantage of the wide \quaia redshift coverage, we measured the growth of structure at $z \simeq 2.7$, achieving one of the highest-redshift constraints on $\sigma_8$ in the literature ($\sigma_8(z=2.7)=0.22\pm0.06$ -- see Equation~\ref{eq:s8z}). Our results, together with other relevant growth constraints from \cite{Miyatake_2022,Farren_2023,GG_2021,DESs8z}, are shown in Figure~ \ref{fig:s8_comp}. Our measurements of $\sigma_8(z)$ are in reasonable agreement (within $1\sigma$) with the evolution predicted by \planck. This result is robust against potential Galactic and extragalactic contamination in the CMB lensing maps, recovering compatible constraints using lensing reconstruction algorithms with varying sensitivity to such contamination. Although our constraints are less precise at $z\lesssim1$ than others found in the literature, this result is non-trivial: it confirms the validity of the standard cosmological paradigm over a broad range of redshifts extending to a largely unexplored regime ($z\gtrsim 2$). \newv{As discussed in Appendix \ref{app:lcdm}, our constraints on $\Lambda$CDM parameters using the newer \quaia catalogue find a value of $\Omega_M$ that is mildly in tension with the best-fit found by \planck ($\sim2\sigma$ higher).}

  We also measure the bias of the quasars and its redshift evolution. We find that, while the overall amplitude of $b(z)$ is comparable to that of previous quasar samples (e.g. eBOSS), it seems to display a milder evolution at high redshifts, more compatible with a scaling $b(z)\propto 1/D(z)$, where $D(z)$ is the linear growth factor. This qualitative behavior seems to be dependent of the cosmological model assumed (or simultaneously constrained), and of the subset of the data considered. The amplitude of the $b(z)$ curve does show some sensitivity to the assumptions made about the cosmological model. Since the \quaia data favors a model with a lower value of $\sigma_8$ than \planck (although not in significant tension with it), when jointly constraining cosmology and $b(z)$, a larger value of the latter is recovered (by $\sim1\sigma$) than that found when fixing all cosmological parameters to the best-fit \planck values. This effect is less relevant when using the {\sf No-TT} $\kappa$ map, given its better agreement with \planck. Despite these shifts, our results are nevertheless largely compatible between different analysis choices.

  Inspired by the slightly different amplitude of the CMB lensing cross-correlation between different reconstruction algorithms (first pointed out in \cite{Alonso_2023}), we investigated the possible presence of foreground contamination on different Planck $\kappa$ maps. If sourced by extragalactic foregrounds, this contamination could affect other cross-correlation analyses. In order to ascertain the redshift range over which this contamination might be most prominent, we divided the total \quaia redshift distribution into six bins and quantified the amplitude of the quasar-$\kappa$ cross-correlation in each of them for 8 different $\kappa$ maps. We find that the effect seems to be localized around redshifts $z \sim 1.7$, where results from different $\kappa$ maps (in particular those using or discarding $TT$ correlations) differ at the $2-3\sigma$ level. \newv{We find no evidence that this tension is the result of Galactic contamination, and source hardening does not ameliorate it either. This leaves extragalactic contamination from a clustered component as perhaps the most likely explanation, although the nature of this contaminant is not entirely clear.} Nevertheless, we find that our constraints on the growth history are not significantly affected by this potential systematic.

  Our analysis highlights the potential of combining CMB lensing data with high-redshift galaxy samples to constrain and validate the $\Lambda$CDM model over the range of cosmic times not directly probed by the CMB or by the low-redshift ($z\lesssim1$) optical surveys that have so far dominated large-scale structure cosmology. In the future, further progress in this regard may be possible thanks to more extensive optical/IR quasar catalogues \cite{DESI,2019BAAS...51g.229S}, samples of Lyman-break galaxies and similar dropout populations \cite{Wilson_2019,Miyatake_2022}, radio continuum surveys, and, potentially, 21cm intensity mapping \cite{2021MNRAS.505.2285B,2021JCAP...12..049S,2019JCAP...07..023C}. In this respect, the cross-correlation with CMB lensing maps will be a vital element to produce reliable cosmological constraints and advances from existing and future experiments, such as the Atacama Cosmology Telescope \cite{ACT:2023kun}, the Simons Observatory \cite{SimonsObs}, and CMB Stage-4 \cite{CMBS4}. In particular, the ability to probe significantly smaller angular scales with higher sensitivity, and over a wide range of frequencies, will make it possible to improve the robustness of these constraints to both Galactic and extragalactic systematics.

\begin{acknowledgments}
  We would like to thank Nestor Arsenov, David Hogg, Andras Kovacs, An\v ze Slosar, Abby Williams and the members of the Quaia team for useful comments and discussions. GP is supported by the INFN project InDark and the ASI/LiteBIRD grant n.2020-9-HH.0. GF acknowledges the support of the European Research Council under the Marie Sk\l{}odowska Curie actions through the Individual Global Fellowship No.~892401 PiCOGAMBAS and of the Simons Foundation. DA acknowledges support from the Beecroft Trust. CGG acknowledges support from the European Research Council Grant No.~693024 and the Beecroft Trust. JC acknowledges support from a SNSF Eccellenza Professorial Fellowship (No. 186879).
 We made extensive use of computational resources at the University of Oxford Department of Physics, funded by the John Fell Oxford University Press Research Fund, and at the Flatiron Institute.
\end{acknowledgments}

\appendix
\section{Comparison with previous results}\label{app:lcdm}

  \begin{table}[h!]
    \begin{center}
      \begin{tabular}{|l|lll|}
        \hline\hline
        Case & $\sigma_8$ & $\Omega_m$ & $S_8$ \\[0.5ex]
        \hline
        $1.\,\,{\rm Old\,\,catalogue,\,\,2\,\,bins}$ & $0.755\pm 0.032$ & $0.346\pm 0.017$ & $0.811\pm 0.040$\\[0.5ex]
        $2.\,\,{\rm New\,\,catalogue,\,\,2\,\,bins}$ & $0.761\pm 0.033$ & $0.359^{+0.017}_{-0.019}$ & $0.832\pm 0.041$\\[0.5ex]
        $3.\,\,{\rm New\,\,catalogue,\,\,Gal.\,\,mask}$ & $0.749\pm 0.042$ & $0.357^{+0.018}_{-0.023}$ & $0.817^{+0.050}_{-0.056}$\\[0.5ex]
        $4.\,\,{\rm New\,\,catalogue,\,\,3\,\,bins}$ & $0.776\pm 0.035$ & $0.352^{+0.017}_{-0.019}$ & $0.841\pm 0.044$\\[0.5ex]
        $5.\,\,{\rm New\,\,catalogue,\,\,3\,\,bin,\,\,Gal.\,\,mask}$ & $0.793\pm 0.042$ & $0.361^{+0.020}_{-0.023}$ & $0.870\pm 0.052$\\[0.5ex]
        $6.\,\,{\rm New\,\,catalogue,\,\,2\,\,bin,\,\,No-TT}$ & $0.809^{+0.050}_{-0.055}$ & $0.361\pm 0.021$ & $0.887\pm 0.069$\\[0.5ex]
        $7.\,\,{\rm New\,\,catalogue,\,\,3\,\,bin,\,\,No-TT}$ & $0.832\pm 0.055$ & $0.356^{+0.021}_{-0.023}$ & $0.906^{+0.065}_{-0.073}$\\[0.5ex]
        \hline\hline
      \end{tabular}
    \end{center}
    \caption{Constraints on $\Omega_M$, $\sigma_8$, and $S_8$ under different analysis setups, comparing the previous and current \quaia catalogues, and quantifying the effects of redshift binning, Galactic masking, and temperature-driven lensing reconstruction.}\label{tab:lcdm}
  \end{table}
  \newv{As described in Section \ref{ssec:data.quaia}, this analysis uses a version of the \quaia sample that is slightly different (updated selection function and better redshift estimates) from that used in the first cosmological analysis of \cite{Alonso_2023}. In addition to this, we have followed different analysis choices, namely the use of 3 redshift bins instead of 2 (to better reconstruct the growth history and isolate the high-redshift contribution), and the study of potential residual Galactic contamination. To quantify the impact of these differences on our analysis, we have derived $\Lambda$CDM from our data and compared them with the results of \cite{Alonso_2023}.}
  \begin{figure}[h!]
      \centering
      \includegraphics[width=0.49\textwidth]{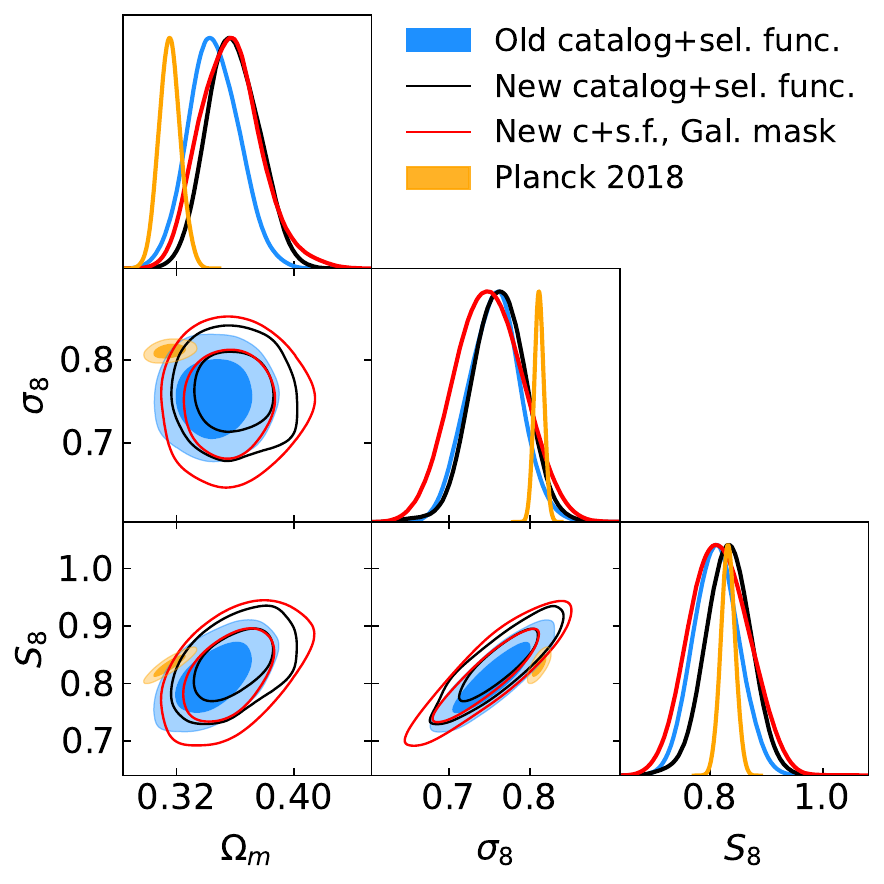}
      \includegraphics[width=0.49\textwidth]{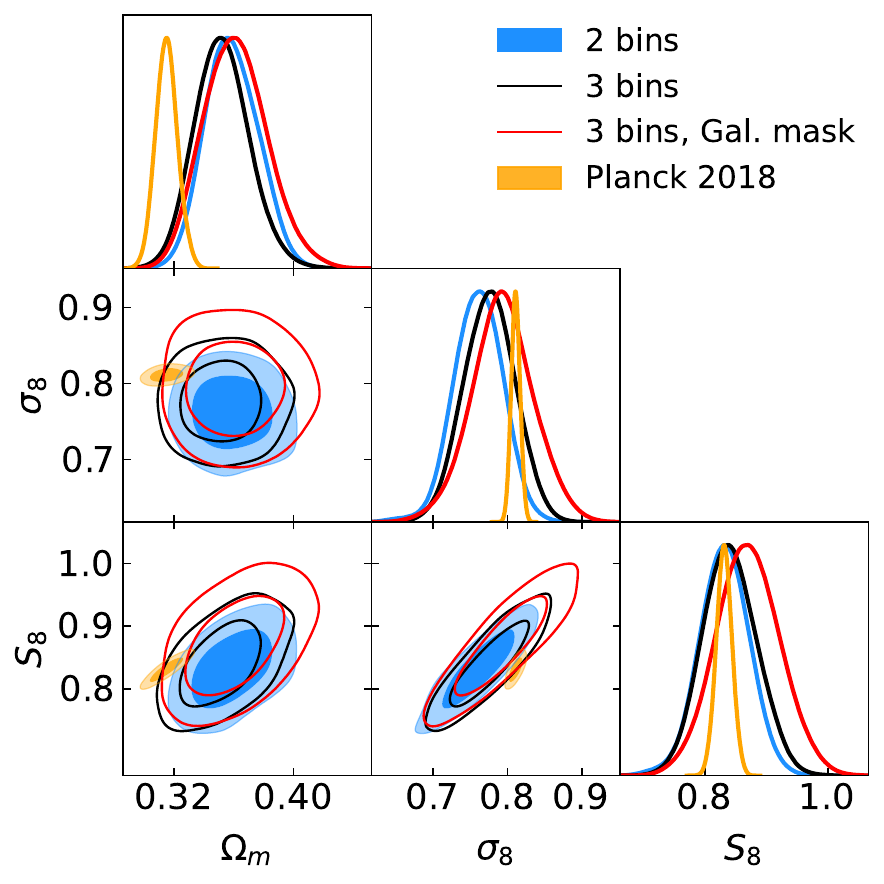}
      \caption{Constraints on $\Lambda$CDM parameters $\Omega_M$, $\sigma_8$, and $S_8$ under different analysis setups, comparing the previous and latest \quaia catalogues, and quantifying the impact of redshift binning and Galactic mask. See main text for details.}
      \label{fig:triangle}
  \end{figure}
  \newv{The left panel of Figure~\ref{fig:triangle} shows the constraints on $\Omega_M$, $\sigma_8$, and $S_8$ obtained from the old catalogue and the new catalogue under the same analysis choices (2 redshift bins), but using different selection functions (blue and black contours). We find identical constraints on $\sigma_8$, although we observe a shift in $0.7\sigma$ upward in $\Omega_M$. Note that the changes in the posteriors are primarily determined by differences in the old and updated selection functions, rather than by the catalogue itself. The observed shift leads to a mild tension with the value of this parameter preferred by \planck (shown in orange in the figure) at the level of $2.3\sigma$. To determine if this might be caused by potential Galactic contamination, we repeat the analysis imposing the $40\%$ Galactic mask from \planck (red contours). This results in virtually the same shift in $\Omega_M$, albeit with $\sim15\%$ larger error bars, commensurate with the loss of sky area. We thus conclude that this mild tension is not caused by Galactic contamination. The right panel of Figure \ref{fig:triangle} then compares the results obtained with the new catalogue using 2 bins (blue) and 3 bins (black), as well as with the 3-bin case using a Galactic mask (red). The use of 3 bins leads to a $0.5\sigma$ upward shift in $\sigma_8$, and a $0.3\sigma_8$ downward shift in $\Omega_M$, reducing the tension in $\Omega_M$ to $1.9\sigma$. Unlike in the 2-bin case, applying the Galactic mask shifts both $\sigma_8$ and $\Omega_M$ upward, although the tension with \planck remains at the same (mild) level. The use of the {\sf No-TT} CMB lensing map (not shown in the figure) does not lead to significant changes in the $\Omega_M$ tension, but leads to a $\sim1\sigma$ upward shift in $\sigma_8$. This is compatible with the results found by \cite{Alonso_2023} with the {\sf Pol-only} map and, as hinted at in Section \ref{ssec:res.fg}, may be caused by residual extragalactic contamination in the $\kappa$ maps using temperature correlations.}

\bibliography{biblio}{}
		
\end{document}